\documentclass[printer]{aa}
\usepackage{natbib}
\bibpunct{(}{)}{;}{a}{}{,} 
\usepackage{graphicx}
\usepackage{tabularx}

\begin{document}

\title{CFBDSIR2149-0403: a 4-7 Jupiter-mass free-floating planet in
  the young moving group AB Doradus ?
\thanks{
}} 
 

\author{
  P. Delorme  \inst{1} 
  J. Gagn\'e \inst{2} 
  L. Malo  \inst{2} 
  C. Reyl\'e \inst{3} 
  E. Artigau \inst{2} 
  L. Albert  \inst{2}
  T. Forveille \inst{1}
  X. Delfosse \inst{1}
  F. Allard \inst{4}
  D. Homeier \inst{4}
}

\offprints{P. Delorme, \email{Philippe.Delorme@obs.ujf-grenoble.fr}. Based on observations obtained with SOFI on the NTT at ESO-La Silla
(run 086.C-0655(A)). Based on observations obtained with X-Shooter on
  VLT-UT2 at ESO-Paranal(run 087.C-0562(A)). Based on observation
  obtained with WIRCAM at CFHT (programs 09AF21,10BF26 and
  11BD86.}

\institute{UJF-Grenoble 1 / CNRS-INSU, Institut de Plan\'etologie et d'Astrophysique de Grenoble (IPAG) UMR 5274, Grenoble, F-38041, France.
    \and D\'epartement de physique and Observatoire du Mont M\'egantic,
  Universit\'e de Montr\'eal, C.P. 6128, Succursale Centre-Ville,
  Montr\'eal, QC H3C 3J7, Canada
  \and Universit\'e de Franche Comt\'{e}, Institut UTINAM CNRS 6213, Observatoire des Sciences de l'Univers THETA de Franche-Comt\'{e}, Observatoire de Besan\c{c}on, BP 1615, 25010 Besan\c{c}on Cedex, France
  \and C.R.A.L. (UMR 5574 CNRS), Ecole Normale Sup\'erieure, 69364 Lyon
  Cedex 07, France
}

\abstract{}
{Using the CFBDSIR wide field survey for brown dwarfs, we
  identified  CFBDSIRJ214947.2-040308.9, a late T dwarf with
  atypically red $J-K_S$ colour.}
{ We obtained an X-Shooter spectra, with signal detectable
  from 0.8$~\mu$m to 2.3$~\mu$m, which confirmed a T7 spectral type with
  an enhanced $K_s$-band flux indicative of a  potentially low-gravity,
  young, object.}
 {  The comparison of our near infrared spectrum with atmosphere
 models, for solar
metallicity, shows that CFBDSIRJ214947.2-040308.9 is probably a
650-750\,K, log\,g=3.75-4.0 substellar object. Using evolution models, this translates
into a planetary mass object, with an age in the 20-200 Myr range. An independent Bayesian
analysis from proper motion measurements results in a 87\% 
   probability that this 
   free-floating planet is a member of the 50-120 Myr old AB Doradus
   moving group, which strengthens the spectroscopic youth diagnosis.}
{ By combining our atmospheric
characterisation with the age and metallicity constraints arising from
the probable membership to the AB Doradus moving group, we find that 
 CFBDSIRJ214947.2-040308.9 is probably a 4-7 Jupiter masses free-floating
 planet with an effective temperature of $\sim$700\,K and a log\,g of
 $\sim$4.0, typical of the late T-type exoplanets that are targeted by
 direct imaging.
 We stress that this object
could be used as a benchmark for understanding the physics of the
similar T-type
exoplanets that will be 
discovered by the upcoming high contrast imagers.}

\date{}


\authorrunning{P. Delorme et al.}
\titlerunning{A 4-7 Jupiter-mass free-floating planet in AB
  Doradus?}
\maketitle

\section{Introduction}
  The Astronomical Union
  definition\footnote{http://www.dtm.ciw.edu/boss/definition.html} that the planetary mass range is below deuterium-burning
  mass \citep[13M$_{Jup}$; ][]{Boss.2003} while
  brown dwarfs and stars populate the mass range above is
  challenged by a string of recent discoveries, notably from  possible
  Isolated Planetary
  Mass Objects \citep[hereafter IPMOs, or equivalently free-floating
    planets in clusters, see for
    instance][]{Zapatero.2002,Burgess.2009,Haisch.2010,Ramirez.2011,Ramirez.2012sub},
  which are more likely formed like 
  stars but reside in the planetary mass range. There are also many
  cases of field brown dwarfs whose lower mass limit is well within
  the official planetary mass range \citep[see for instance
][]{Knapp.2004,Burgasser.2006,Cruz.2009,Lucas.2010,Burningham.2011a,
Liu.2011,Luhman.2011,Cushing.2011,Albert.2011}.  However, in all these cases
  there are significant uncertainties on the actual masses of these
  possible free-floating planets, mostly because of the
  age/mass/luminosity degeneracy that affects the determination of the
  physical parameters of substellar objects. This degeneracy can be
  lifted when the age of the source can be constrained
  independently, usually through cluster or association
  membership. There is however no isolated object that combine such an
  undisputed age constraint with spectroscopic low gravity signatures that would be compatible with a planetary mass. There is
  nonetheless strong evidence that IPMOs do exist, at least since the
  discovery of 
  2M1207B by \citet{Chauvin.2004} has established the existence of a $\sim$5M$_{Jup}$
  companion around 
a $\sim$25M$_{Jup}$ object that would be almost impossible
  to form through planetary formation mechanisms. This means stellar
  formation processes such as cloud fragmentation \citep[see for
    instance][]{Bate.2009} or disk fragmentation \citep[see for
    instance][]{Stamatellos.2011}, can form planetary mass
  objects. \citet{Ramirez.2012sub} has recently identified a population
  of IPMOs in the $\sigma$-Orionis cluster, and hinted that they could
  be about as numerous as deuterium-burning brown-dwarfs. This would
  indicate that there is a significant population of overlooked IPMOs
  in the solar vicinity, both in the field and in young
  moving groups and clusters. Another source of IPMOs could be ejected
  planets \citep[e.g.][]{Veras.2012,Moeckel.2012}, 
  since massive planets such as HR8799bcde \citep{Marois.2008,Marois.2010HR8799e}, if ejected from their host
  star, would look like regular field T dwarfs after a few hundred
  Myr. Another strong evidence that IPMOs exist is the detection of a few
  free-floating planets by gravitational lensing by
  \citet{Sumi.2011,Strigari.2012}, though these objects -or at least a fraction of them- could also be
  regular planets orbiting at sufficiently large separation from their host
  star that the latter is not detectable in the lensing event.\\

The detection of IPMOs can therefore provide constraints on ejection
scenarii and on the low-mass end of the stellar mass function, though
these constraints will not be independent since it is observationally
challenging to imagine a way to discriminate between a
5M$_{Jup}$ ejected planet and a brown dwarf of the same mass. However, the spectral energy distribution of these
   isolated objects will provide useful information on the
   substellar evolution and substellar atmosphere models, especially
   if their age is known, and regardless of their formation
   mechanisms. These constraints are especially valuable 
   because the spectral energy distribution of an IPMO is expected to be
    identical to the 
   spectral energy distribution of planets with similar masses
   orbiting at large separation -hence with negligible irradiation- from
   their host stars. In this light, such free-floating planets could
   serve as benchmarks for the design and operation of the direct
   imaging surveys for exoplanets, notably with the upcoming new
   generations planets finder instruments such as SPHERE \citep{Beuzit.2008},
   GPI \citep{Graham.2007}, or HiCIAOH \citep{Hodapp.2008}. Young IPMOs of a few Jupiter masses would be interesting
   analogs of the exoplanets these instruments will be able to
   detect. Since they are not affected by the glare of a host star, it
   is comparatively easy to obtain 
   relatively high signal to noise, moderate resolution spectroscopic
   information so that they can serve as prototypes to understand the
   physics of 
   massive exoplanets atmospheres.\\

 We present in section 2  the detection of a low-gravity T
 dwarf, CFBDSIRJ214947.2-040308.9, hereafter CFBDSIR2149, that 
 is probably a 4-7M$_{Jup}$ free-floating planet.  In section 3, we
 describe our spectroscopic data reduction and present the full
 Near-InfraRed (NIR) spectrum of this object, highlighting
 its low gravity features. 
In section 4 we
 discuss its likely membership to the young moving group
 AB Doradus (hereafter ABDMG). In section 5 we analyse its spectrum in the light of the
 age and metallicity constraints that would be brought by its probable
 ABDMG membership, and derive its fundamental physical
 parameters. Finally we discuss some of the implications of this discovery.


\section{Photometry of CFBDSIR2149} \label{observations}  
  \subsection{Discovery and identification as a late T dwarf with
    $K_s$-band excess}
The Canada-France
  Brown Dwarf Survey InfraRed \citep[CFBDSIR;][]{Delorme.2010} is a
  NIR coverage of the CFHTLS/CFBDS 
  \citep[see][]{Delorme.2008b} fields with available deep $z'$ band
  images. It covers 335 square degrees with  $J$-band WIRCam
  \citep{Puget.2004} images.  Cool brown dwarfs candidates are
  identified from their very red $z'-J$ colour (see Fig. \ref{colspt}) and 
  CFBDSIR2149 was identified as such by the standard CFBDSIR analysis pipeline 
  \citep[see ][]{Delorme.2010}. 
 After analysing and cross-matching a stack of
  two 45s-long WIRCam $J$-band exposure from the CFBDSIR acquired on
  August 13th and 14th, 2009, with the corresponding CFBDS-RCS2 
  \citep{Yee.2007,Delorme.2008b}   
  360s $z'$ exposure, we highlighted CFBDSIR2149 as a $z'$
  dropout with $z'-J>$3.8. 

\begin{figure}
 \caption{\label{finding_chart}Finding chart for CFBDSIR2149 ($Js$-band
   NTT Sofi image). East is left and north is up.}
\includegraphics[width=8cm]{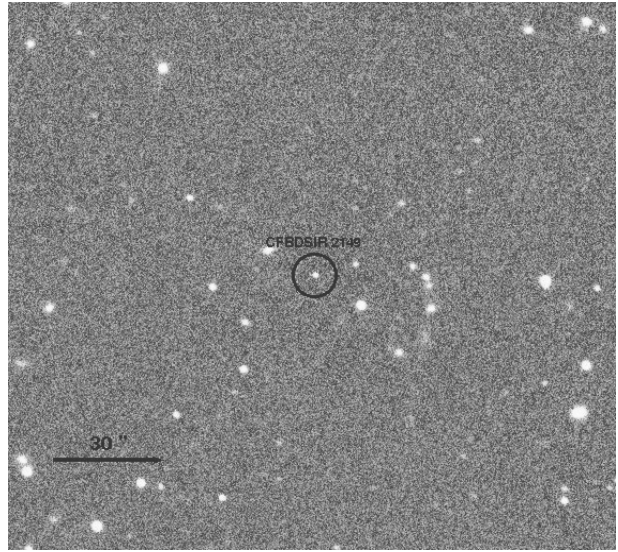}
\end{figure}

 This promising candidate was confirmed by
  NIR follow-up observation at the ESO-NTT 
  telescope (run 086.C-0655(A)) on 2010, September 24, with a NTT-Sofi \citep{Moorwood.1998}
  $J$-band detection (See Fig. \ref{finding_chart})
  ensuring this candidate was a very cool brown dwarf and not
  a transient source like an extragalactic supernovae or an asteroid
  that could have caused our initial $J$-band detection and accounted
  for the $z'$-band non-detection. The real-time analysis of the NTT
  $J$-band data prompted further observations of
  this cool brown dwarf in $H$ and $K_s$, during the following
  night. The resulting 
  very blue $J-H=-$0.5 confirmed it as a very late T dwarf while the
  neutral to red $J-K_s \sim$0 highlighted it as a peculiar $K_s$-band
  flux enhanced late T dwarf (see \ref{colspt}). Note that throughout this work, $YJHK/K_s$
  magnitudes are given in the $Vega$ system while $z'$ mags
  are in the $AB$ system \citep{Fukugita.1996}.

\subsection{Near-infrared imaging: reduction and analysis}
   In addition to this SOFI photometric follow-up, we obtained
   higher signal to noise photometric data in $Y$, $J$ and  $K_s$ from
   CFHT Director Discretionary Time, in December
   2011, as well as $CH_{4on}$ imaging acquired in
   September 2010. This WIRCam 
   data, using the MKO photometric system, is deeper and offers a
   larger time-base to derive the proper
   motion of CFBDSIR2149 using the WIRCam August 2009 detection images as a
   first epoch. Since WIRCam has a larger field of view than
   SOFI it also allows to use more stars for an accurate photometric and
   astrometric calibration. 

 Both SOFI and WIRCam observations used
 a  standard dithering pattern to allow the construction of a sky
 frame and were reduced and analysed using the same home-made 
 pipeline. For the WIRCam observations, we used one (out of 4) $10\arcmin \times
 10\arcmin$ chips, on which the target was centred. For each filter, flat
 fielding and bad pixel removal were carried out using 
 ESO-\textit{eclipse} software package \citep{Devillard.2001}. A sky frame,
 constructed by median-combining the dithered raw exposures, was
 subtracted to each exposure. The resulting reduced individual
 exposures were cross matched using \textit{Scamp} \citep{Bertin.2006} and combined
 with \textit{Swarp} \citep{Bertin.2010}, using the inverse of each image
 background noise as weight. This weighting particularly improved the
 signal to noise of
 the WIRCam images which were acquired with a rapidly evolving
 airmass, while the target was setting.
 The absolute astrometric and photometric calibrations in $J, H$ and $K_s$ were
 carried out 
 using the 2MASS point source catalog \citep{Cutri.2003} as a reference, with 3 valid
 references on the SOFI field of vue and 15 on the WIRCam chip field
 of view. For the photometric calibration of $z'$, $Y$ and $CH_{4on}$ data we
 used the CFHT-provided zero points and absorption values.

\begin{figure*}
 \caption{\label{colspt}Near-IR MKO colours of CFBDSIR2149 (Red square) compared to
   known field T dwarfs. The arrows in $z'-J$ colours indicates this
   colour is a lower limit since we have no $z'$-band detection. Open squares are 
CFBDS T dwarfs \citet{Albert.2011}, the plus signs are 111 dwarfs from \citet{Knapp.2004},
crosses are 73 dwarfs from \citet{Chiu.2006} while open triangles are late T
dwarfs from \citet{Burningham.2008,Burningham.2009,Burningham.2010,Delorme.2010,Lucas.2010}.}
\includegraphics[width=16cm]{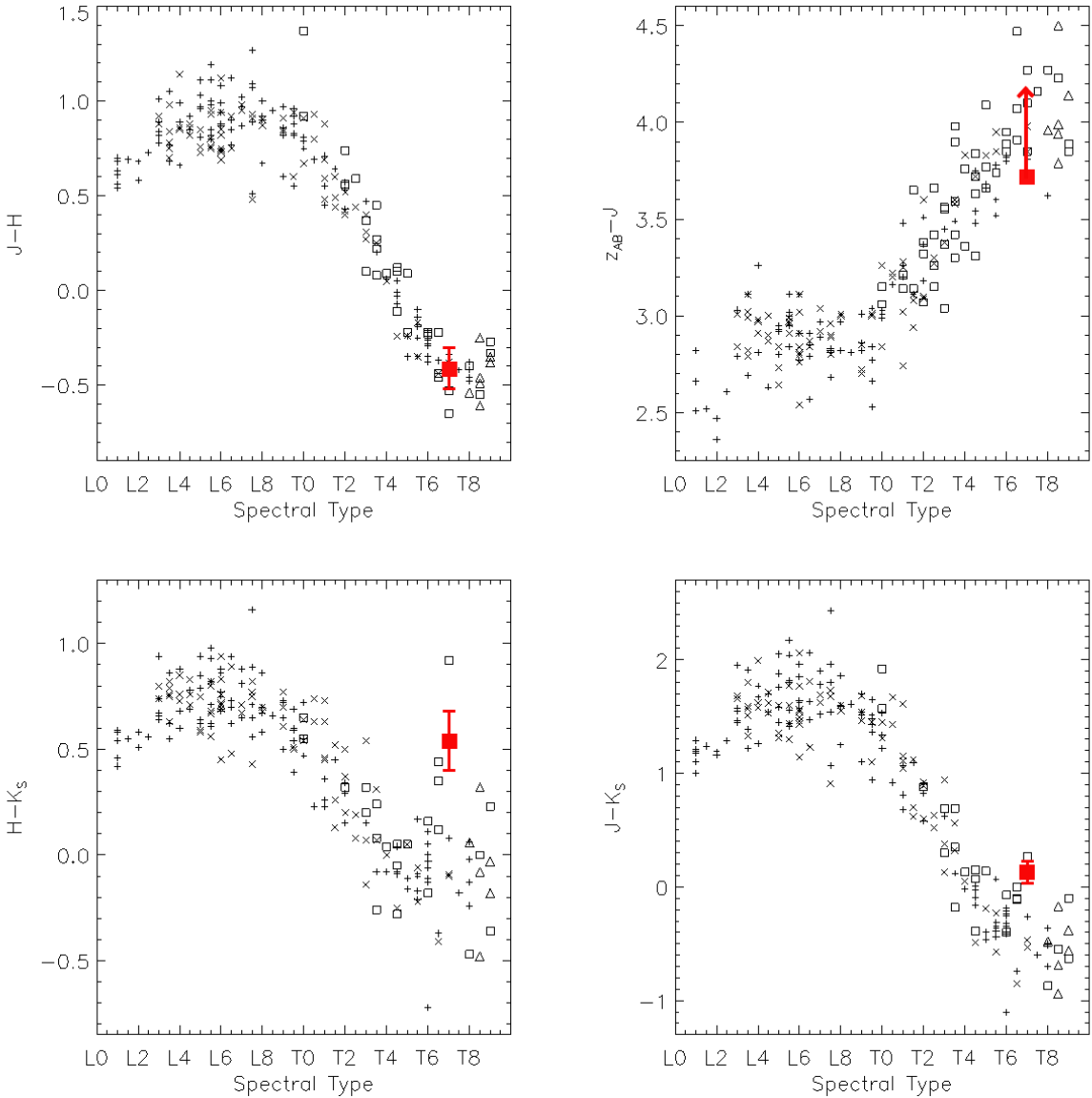}
\end{figure*}

  We extracted the photometry and astrometry of CFBDSIR2149 and of the reference
  stars by Point Spread Function (PSF) fitting using
  \textit{Sextractor}, with a  spatially variable PSF
  model built from each science image using \textit{PSFex} \citep{Bertin.1996,Bertin.2006,Bertin.2012}. The
  resulting photometry is shown on Table \ref{phot}. The $H-CH_{4on}$
  of 0.9, tracing the methane absorption bands around 1.6$~\mu m$ is
  typical of a T7$\pm$0.5 brown dwarf
  (calibration by L. Albert, private communication), while the red $J-K_s$ and very red $H-K_s$ (see
  Fig. \ref{colspt}) indicate a
  weak Collision Induced Absorption of $H_2$ (CIA),
  resulting in an enhanced $K$-band flux \citep[see][ for
    instance]{Knapp.2004}. The weak CIA would be caused by a
  lower than usual pressure in the photosphere, either due to low
  gravity, high metallicity or a combination of both. We caution that
  such colour diagnosis can be misleading, at least for objects in the
  L/T transition where cloud coverage effects can blur the conclusions
  \citep[see the unusual blue and red $J-K$ in the T2.5/T4
    binary of][]{Artigau.2011}. However the later spectral type of
  CFBDSIR2149 (T7) places it in a temperature range where models, and
  the associated colour diagnosis, are
  usually more reliable, mainly because most clouds have condensated at
  such low temperatures. This also make the red $J-K$ association with
  low gravity/high metallicity more robust for late T dwarfs such as
  CFBDSIR2149 than it is for the red, also probably low gravity, L dwarfs identified by \citet{Cruz.2009,Allers.2010,Faherty.2012sub}.

\begin{table*}
\begin{tabular}{c|c|c|c|c|c|c} \hline \hline 
    & $z'_{ab}$  & $Y$ &  $J$ &  $H $ &  $K_s$  & $CH_{4on}$  \\ \hline
CFHT  & $>23.2$ (360s) & 20.83$\pm$0.09 (936s)& 19.48$\pm$0.04 (540s) &  -  &
19.35$\pm$0.09 (880s) &20.7$\pm$0.25 (900s)\\
 Date  & 05/07/2006  & 18,26/12/2011  & 18,26/12/2011  &  -  & 18,26/12/2011 &  29/09/2010 \\ \hline 
NTT &  -  & -  & 19.48$\pm$0.04 (720s) &  19.89$\pm$0.11 (600s) &
19.54$\pm$0.14 (1200s) & - \\
Date &  -  & -  & 23/09/2010  &  24/09/2010  &  24/09/2010  & -   \\ \hline 
\end{tabular}
\caption{Photometry of CFBDSIR2149 with NTT and CFHT. The
  corresponding exposure time is indicated between parentheses.  \label{phot}
}\end{table*}

\subsubsection{WISE data}
  We searched the WISE \citep{Wright.2010} all-sky release catalog for a
  mid-infrared couterpart of CFBDSIR2149. There is no signal
  related to our target in W3 and W4 channels, but we found a very faint object
  at its exact position in W2 and slightly offset in W1. The
  catalog-provided photometry for this couterpart gives W1=17.4$\pm$0.47
  and  W2=15.99$\pm$ 0.37. Though this mid-infrared signal
  to noise is extremely low, the resulting $H-W2$=3.9$\pm$0.4 colour
  is consistent with CFBDSIR2149 being a late T dwarf \citep[see
    Figure 1 of ][]{Mainzer.2011}.

\section{Spectroscopy of CFBDSIR2149}
\subsection{Spectroscopic follow-up and reduction}
  Given the faintness ($J=19.3$) of CFBDSIR2149, it was not 
  observed in spectroscopy with Sofi at NTT and was therefore put in
  the queue of our 087.C-0562 ESO-VLT X-Shooter observations as a very
  high priority target. It was observed on September, 5 and September
  27, 2011, in two ESO observing block (hereafter OB) achieving a
  total exposure time on target of 5850\,s, split in 4 A-B nods on slit
  of 2x732\,s each. We used 0.9$\arcsec$ slits for both visible and NIR
  arms.

 The spectra was reduced using the latest ESO X-Shooter
  pipeline \citep{Modigliani.2010}, which produced a 2-dimensions,
  curvature corrected, spectrum of the NIR arm of
  X-Shooter from 0.99 to 2.5$~\mu$m and of the visible arm, from 0.6 to 1.02$~\mu$m for each OB. No signal
  was retrieved for wavelengths shorter than $\sim$0.8$~\mu$m, but a low
  signal-to-noise ratio (SNR) spectrum of the optical far-red was
  recovered between 0.8 and 1.0$~\mu$m. 
 The trace was extracted using our own IDL
  procedures, using Gaussian boxes in the spatial dimension all along
  the spectral direction. The noise spectrum was obtained by measuring  the dispersion along 10 spectral pixels on a noise trace obtained by
  subtracting the science trace by itself shifted of 1 pixel. Since
  the shift is much smaller than the full spectral resolution ( 4.2
  pixels in the NIR and 6.0 in the visible), this effectively removes
  the science spectrum, but keeps the information on the actual
  background and photon noise on the science trace.

  The two resulting 1-D spectra of each OB were then divided by the
  spectrum of telluric standard stars observed just after or just before each
  OB and reduced and extracted using the same pipeline as the science
  OBs. The two spectra corrected from telluric absorption were then
  combined through a weighted average using the inverse variance as
  weight to construct the science spectrum. A
  noise-weighted average significantly improved the signal-to-noise
  ratio 
  since the quality of data obtained on September 5th was noticeably
  better than that obtained on September 27th. Because the resulting
  spectrum at R$\sim$5300 has a low SNR and most of
  the late T physical parameters exploration is carried out using lower
  resolution spectra, we made a sliding noise-weighted average on 
  25, 100  and 200 pixels in the spectral dimension, producing 3 spectra,
  at a resolution of respectively R=900, 225 and 113 in the
  near-infrared. The visible spectra was similarly binned on
  400 pixels, with a corresponding resolution R=132. The
  noise-weighted average in the 
  spectral dimension makes use of the
  full resolution of X-Shooter to down-weight  the narrow
  wavelength ranges affected by OH telluric emission lines, thus improving
  the SNR with respect to a regular average or, more importantly, to lower resolution observations. 

The same reduction and extraction procedures were used for the NIR and
visible arms of X-Shooter, but the SNR in the small common
  wavelength interval between the 2 arms is low (about 1), which makes difficult
 to rescale the visible data to the NIR data scale. Since we have no
 $z'$ detection of CFBDSIR2149, we cannot calibrate the visible
 spectrum on photometry.  We therefore
  caution that this crude rescaling is probably 
  inaccurate. Therefore, when NIR and visible spectra are shown together, their
  relative intensity before and after 1$~\mu m$ is not reliable.

  We checked the flux homogeneity of this large wavelength coverage
  spectral data by calibrating it on our existing WIRCAM and NTT
  photometry (see Table\ref{phot}).  We synthetized the uncalibrated
  science spectrum colours by integrating them on WIRCam global
  transmission, 
  including filter, instrument and telescope transmission and the
  detector quantum efficiency \citep[see section 2.2 of ][for details]{Delorme.2008b} and determined a
  scaling factor for each of the broadband filter range so that the
   $YJHK_s$ Vega magnitudes derived for CFBDSIR1458AB spectrum would
  match those observed in broadband photometry. The resulting calibration
  factors are summarised in Table \ref{spec_phot}, showing that photometry
  and spectrophotometry agree within 1$\sigma$.


  As shown on Table \ref{spec_phot}, we also derived
  spectrophotometric $CH_{4on}$ and $CH_{4off}$ magnitudes from the
  spectra. Since we don't have a parallax for
  CFBDSIR2149 yet, we can only derive colours from the spectra. We had to
  anchor these colours to the $J$-band photometry to obtain
  the spectrophotometric  $CH_{4on}~CH_{4off}$
  magnitudes in the WIRCam photometric system. 

  We used this X-Shooter spectrum to derive the spectral
  indices defined in \citet{Burgasser.2006,Warren.2007,Delorme.2008a}
  that trace the strength of several molecular absorption features
  typical of T dwarfs. As shown on Table \ref{indices}, the
  atmospheric features are typical of a T7-T7.5 dwarf, with a
  significantly enhanced $K/J$ index, telltale of a weak
  CIA \citep[though greenhouse effect could participate
  to $K$-band flux enhancement, see][]{Allard.2012sub}, and therefore
  of a low pressure photosphere
  \citep{Leggett.2002,Burgasser.2004,Golimowski.2004,Knapp.2004,Burgasser.2006}.
\citet{Hiranaka.2012} propose an alternative explanation for the
similarly red spectral energy distribution of some peculiar L dwarfs,
which could be caused by a thin dust layer above the
photosphere. Since most of the dust is condensated in late T dwarfs
photospheres, this alternative hypothesis is much weaker for objects
as cool as CFBDSIR2149, making a lower than usual pressure in
the photosphere the most likely hypothesis to explain the red $J-Ks$
colour of this object.
 Such
  a low pressure can be
  the sign of a young, low-mass and therefore low-gravity object
  and/or of a more opaque, higher altitude photosphere
  typical of a high metallicity object. \\
  
\begin{table*}
\caption{Value of the NIR spectral indices from
  \citet{Burgasser.2006,Warren.2007,Delorme.2008a} for some of the
  latest known brown dwarfs. We calculated the values the other brown
  dwarfs using spectra from
  \citet{Burgasser.2006,Delorme.2008a,Burningham.2009,Burningham.2010}.
\label{indices}}     
\begin{tabular}{|l|c|c|c|c|c|c|c|c|c|} \hline   
  Object       &Sp. Type&   H$_2$O-J &W$_j$ &      CH$_4$-J &
  H$_2$O-H  &      CH$_4$-H   &      NH$_3$-H& 	 CH$_4$-K & K/J \\ \hline
{\bf CFBDSIR2149} &  T7/T7.5  & 0.067 & 0.404& 0.198
&  0.222 & 0.138  &  0.706  &
0.133  & 0.199 \\ 
      &  & $\pm$0.003  & $\pm$0.004 &$\pm$0.004  &$\pm$0.007
&$\pm$0.006  & $\pm$0.015  & $\pm$0.023  &  $\pm$0.003 \\ 
     &    & T7.5 &  T7 & T8 & T7 & T7.5 & -  & T6.5&  - \\ \hline
SDSS1504+10 &  T7  & 0.082  & 0.416  &  0.342  & 0.241  &  0.184  &
0.668 & 0.126 & 0.132   \\  \hline
Gl570D &  T7.5  &  0.059 & 0.330  & 0.208   &0.206  & 0.142   & 0.662  & 0.074 & 0.081  \\ \hline
2M0415 & T8 &  0.030 & 0.310  & 0.172   & 0.172  & 0.106   & 0.618  & 0.067 & 0.133   \\ \hline
Ross458C &  T8+  & 0.007  & 0.269  & 0.202   & 0.219 &  0.107  & 0.701 & 0.082 &  0.192  \\  \hline
Wolf940B&T8+  & 0.030&  0.272 & 0.030 & 0.141   &    0.091   &
0.537& 0.073 &0.111 \\ \hline 

\end{tabular}
\end{table*}

\begin{figure*}
 \caption{Comparison of CFBDSIR2149 full spectrum at R=225 with BT-Settl models of
  varying effective temperature (left) and gravity (right). Last
  row shows the models agreeing best with $J,H,K$ data for field
  gravity(left, log\,g=4.5 and T$_{eff}$=800\,K) and free-floating planet
  gravity (right, log\,g=3.75 and T$_{eff}$=650\,K)  \label{spectraVsmodelsTeff}}
\begin{tabular}{c|c} 
\includegraphics[width=8cm,angle=0.]{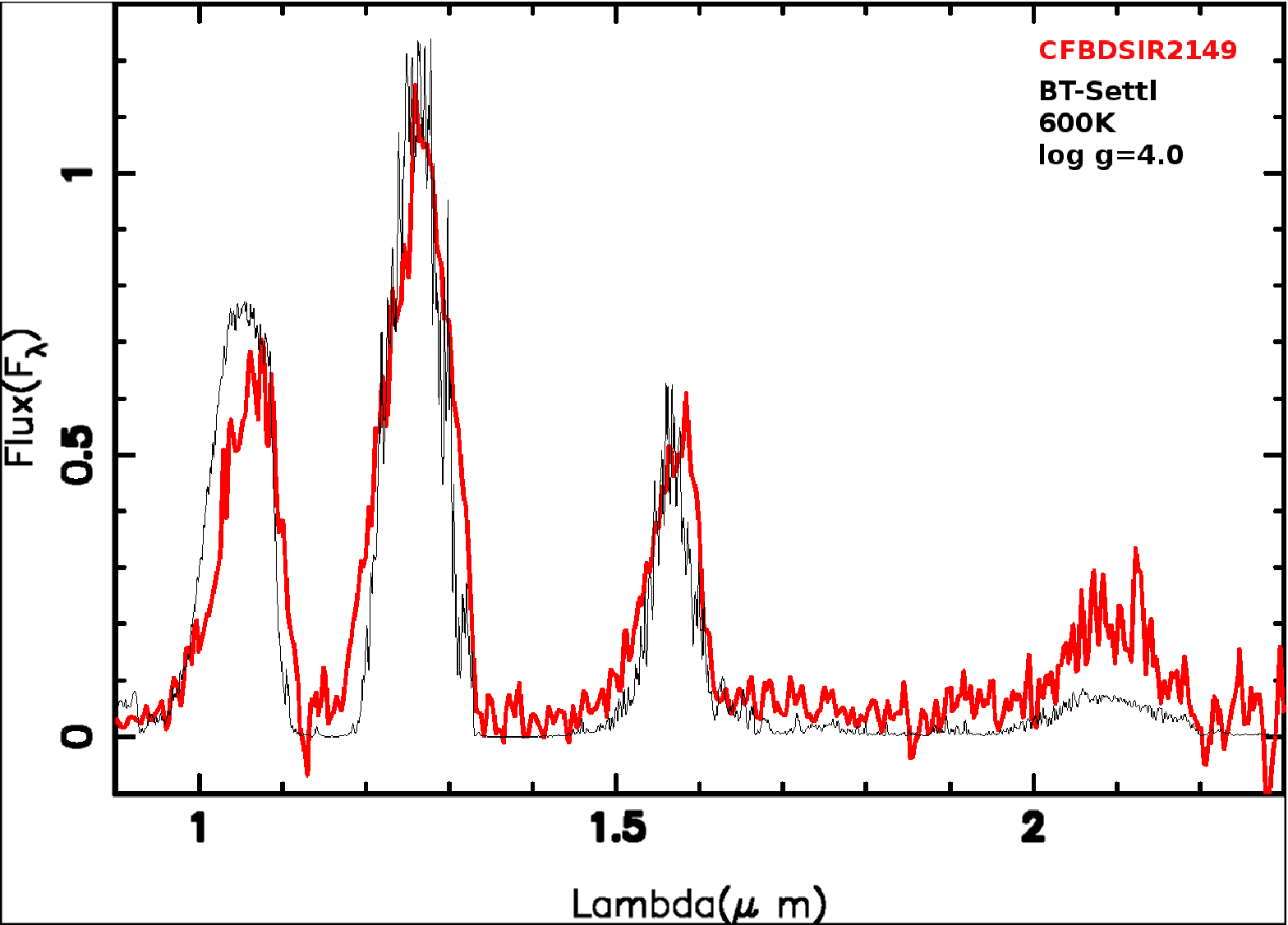} & \includegraphics[width=8cm,angle=0.]{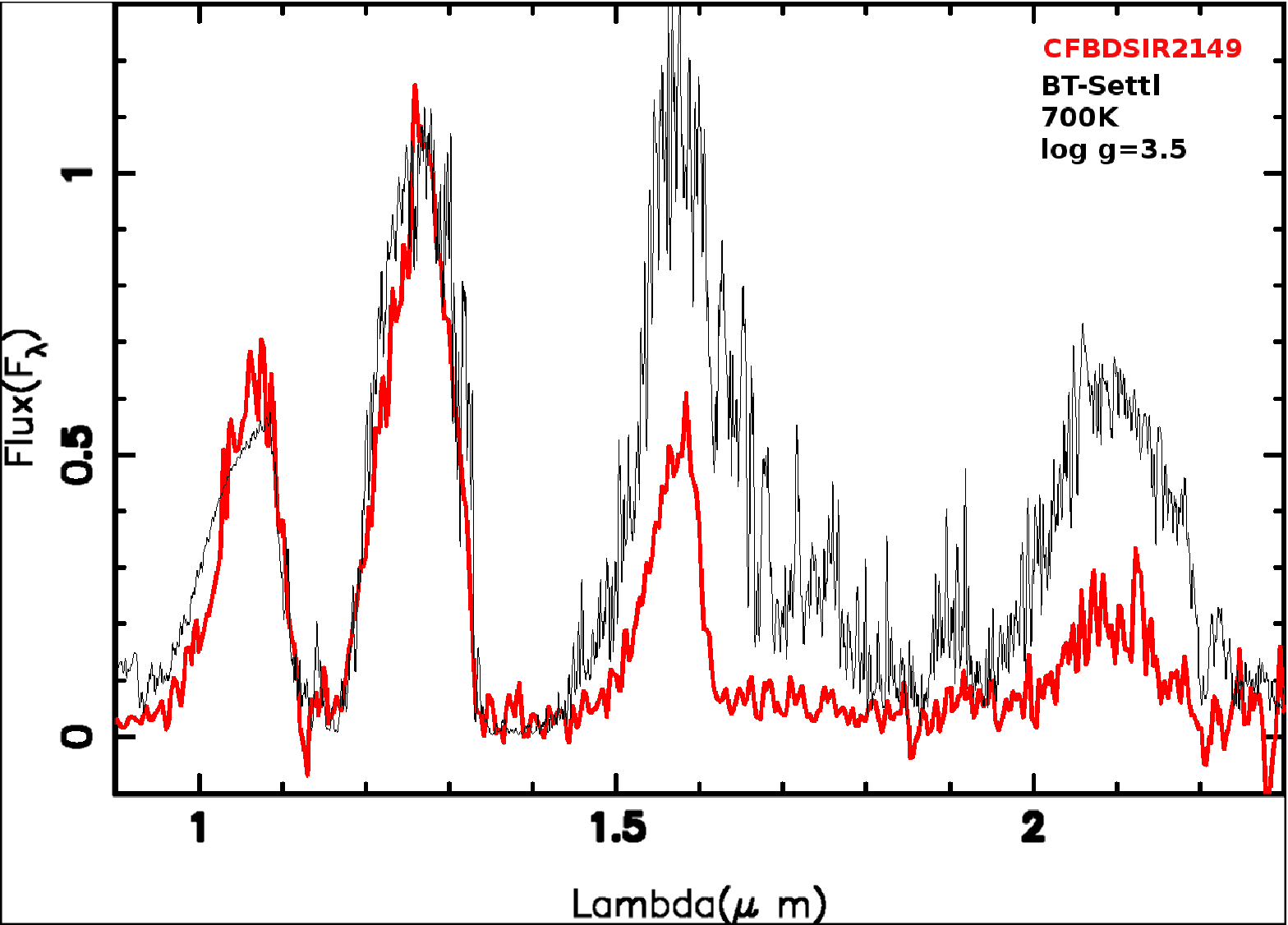}\\ 
\includegraphics[width=8cm,angle=0.]{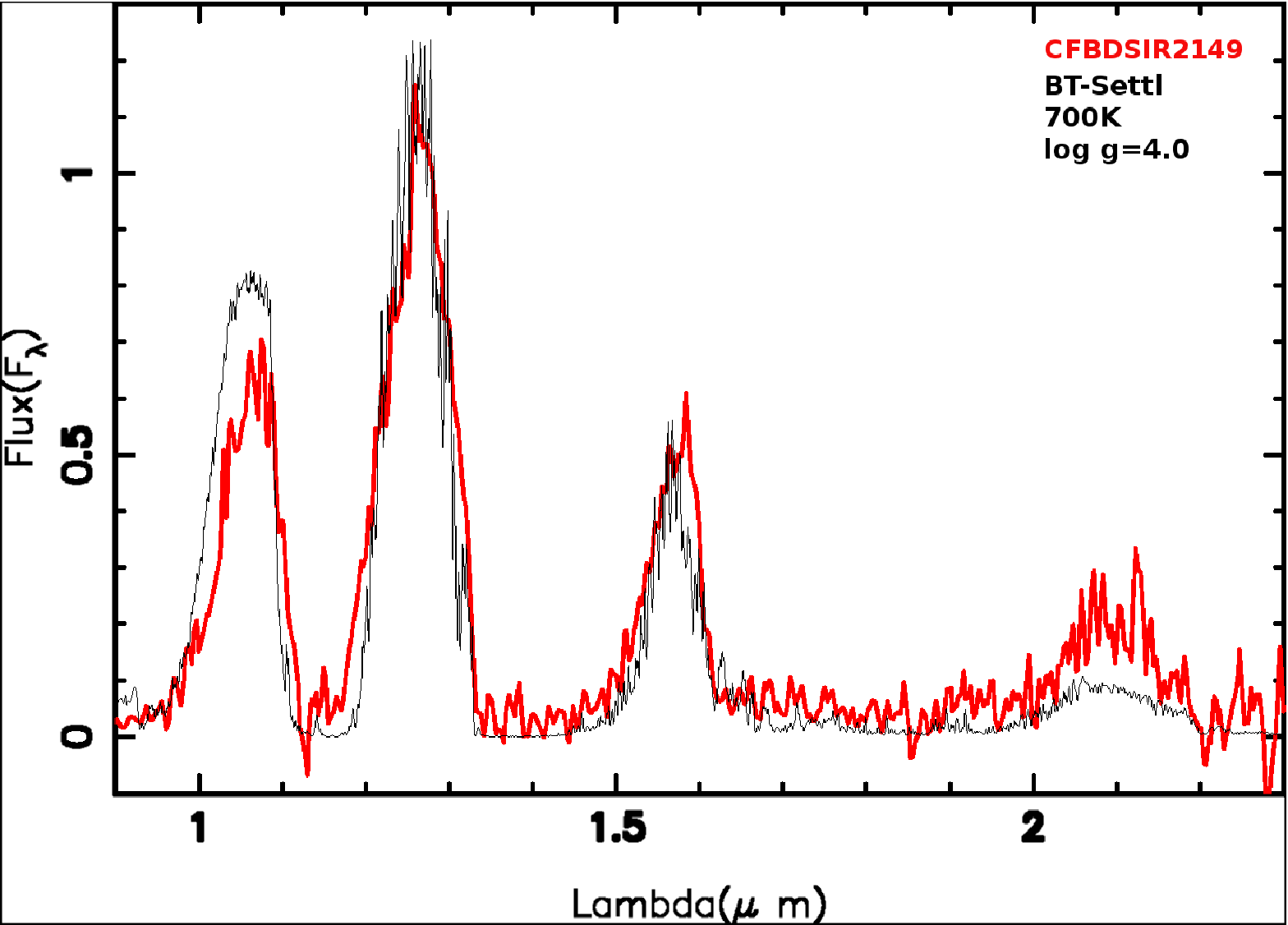} & \includegraphics[width=8cm,angle=0.]{CFBDSIR2149_Vs_lte007-4_0.ps}\\
\includegraphics[width=8cm,angle=0.]{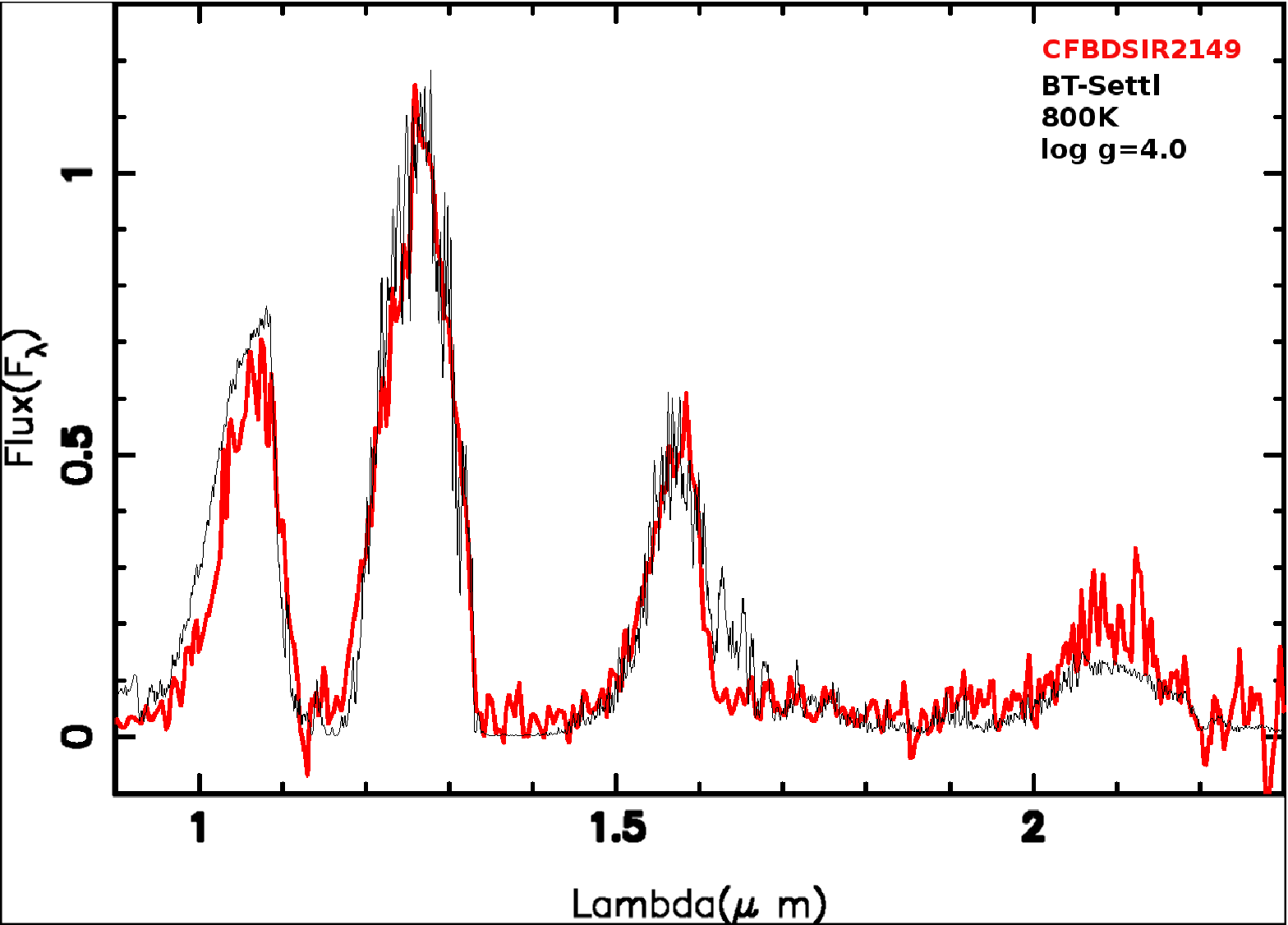}
&
\includegraphics[width=8cm,angle=0.]{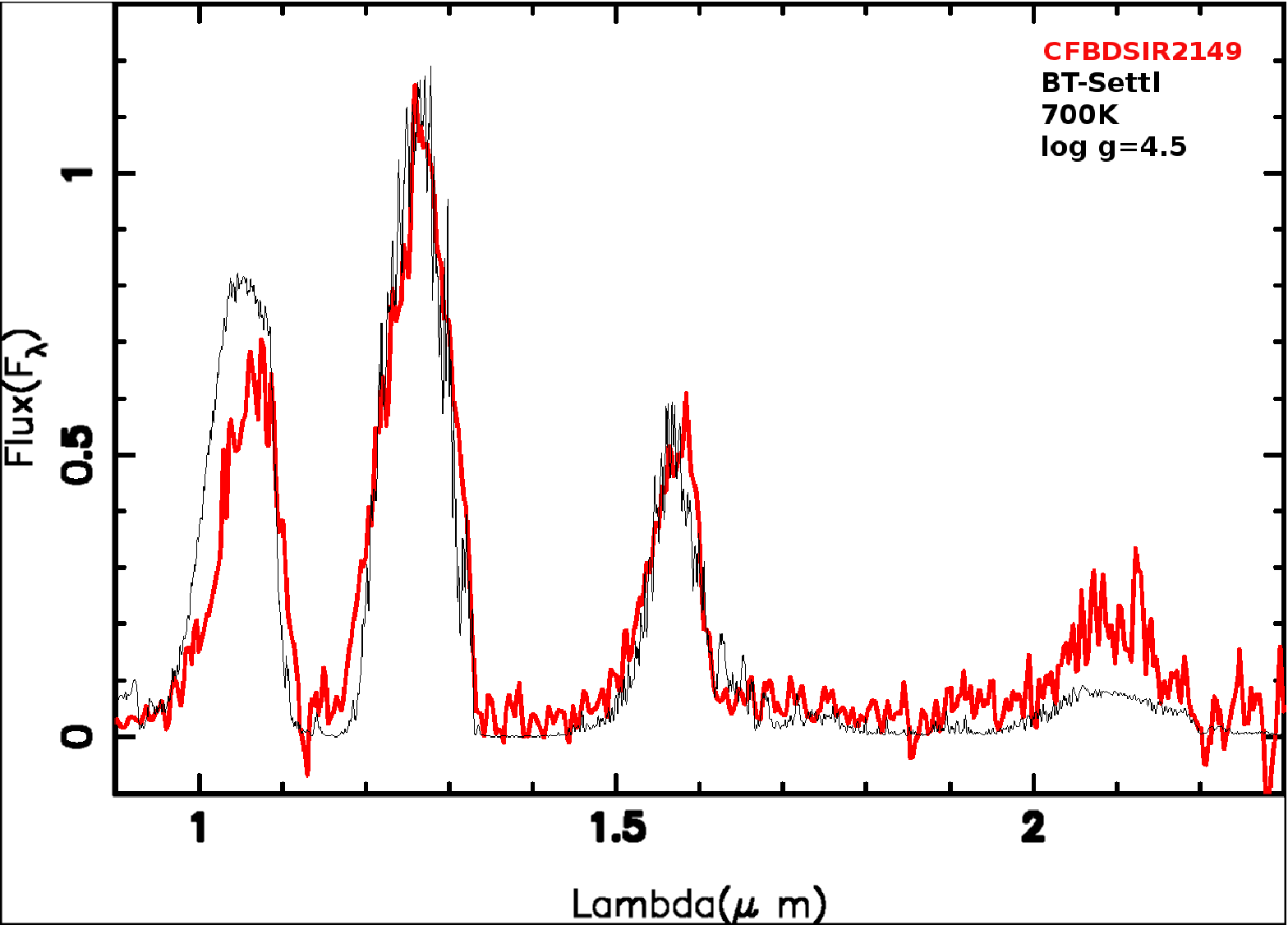}\\ \hline
\hline
 & \\
\includegraphics[width=8cm,angle=0.]{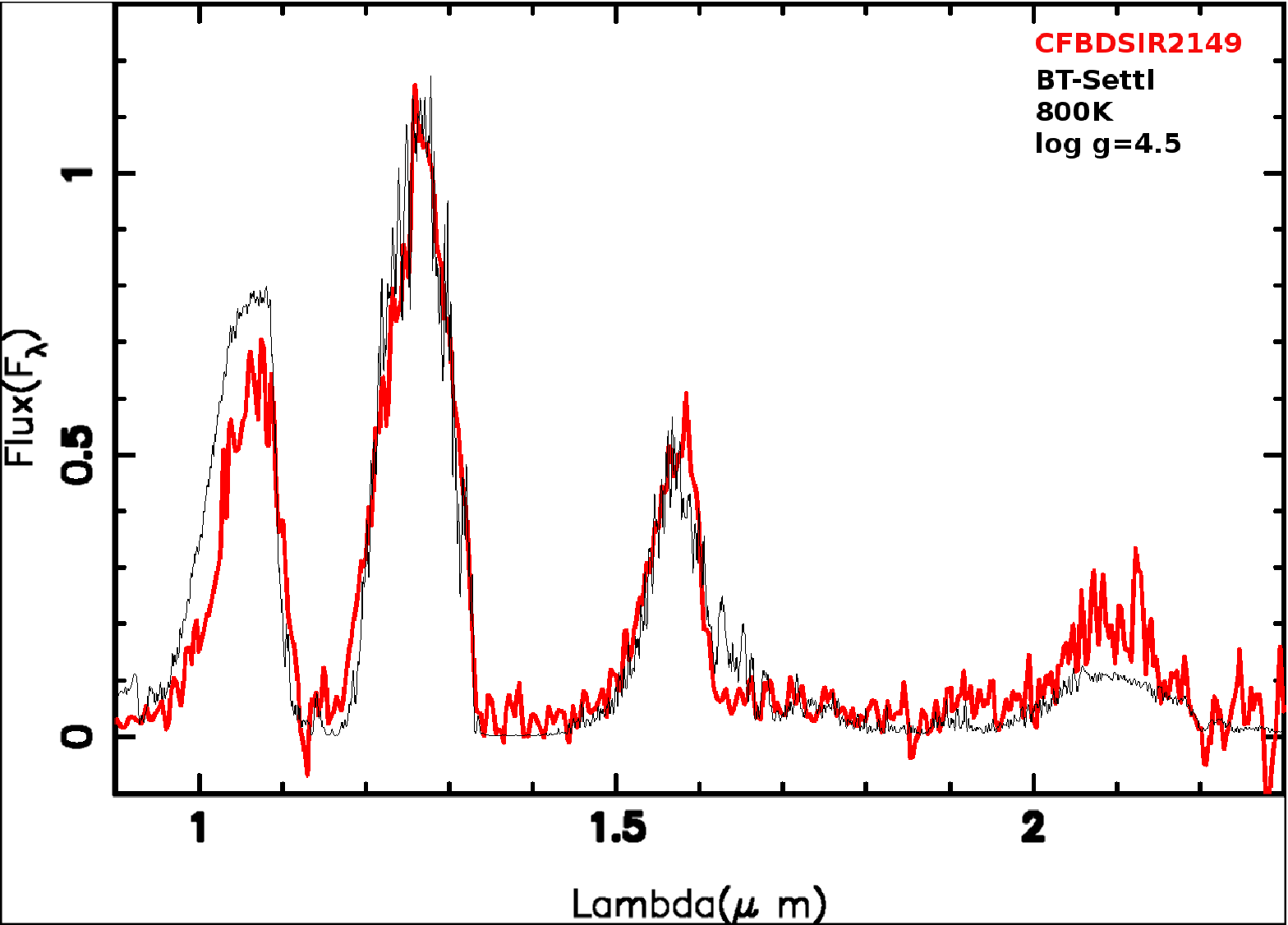} & \includegraphics[width=8cm,angle=0.]{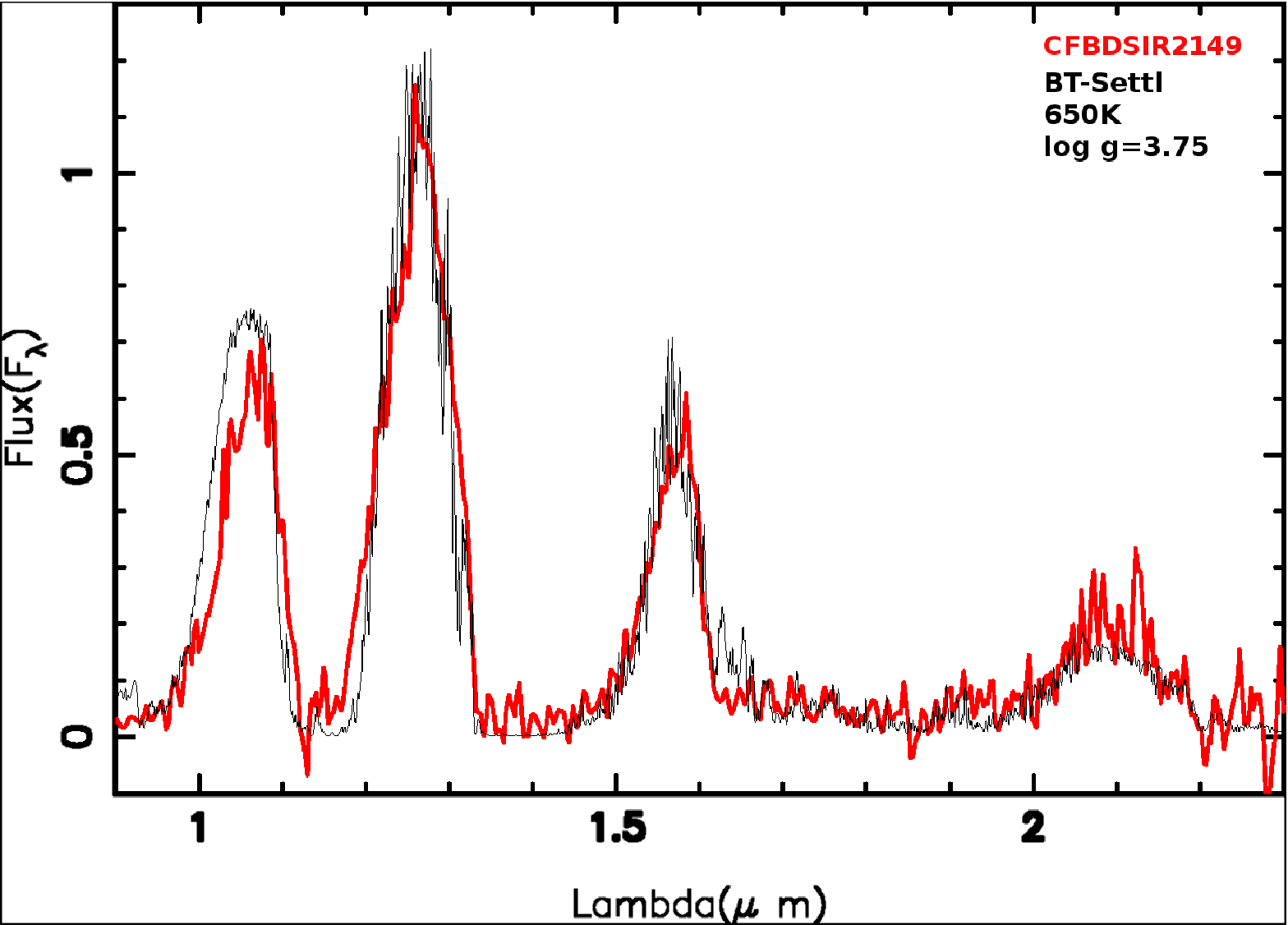}\\
\end{tabular}
\end{figure*}

Figure \ref{spectraVsmodelsTeff} presents a simple comparison of
CFBDSIR2149 spectrum with BT-Settl atmosphere models from
\citet{Allard.2012} at different effective temperatures and
gravity. Though we defer a more exhaustive analysis of the spectrum to
a dedicated section, this first glance at our target's spectrum shows
the following:
\begin{itemize}
\item Models at moderate gravity (log\,g$\geq$4.5) typical of relatively
  young thin disc objects aged 0.5-2Gyr old  \citep[from stellar
    evolution models 
  of ][]{Baraffe.2003} cannot reproduce both the
  strong absorption bands in $H$-band and the enhanced flux in
  $K$-band. Higher gravity models (log\,g $>$5.0) are even more
  discrepant from the observed spectrum.
\item Models at very low gravity (log\,g$\sim$3.5) typical of very young
  objects (1-20Myr) predict a $K$-band (and $H$-band) flux enhancement
  much stronger 
  than what we observe for CFBDSIR2149.
\item Models at low gravity (log\,g$\sim$3.75-4.0) typical of
  intermediate age
  objects (20-200Myr) produce a spectral energy distribution
  relatively close to the observed spectrum.
\item Models at temperature higher than 800\,K significantly
  underestimate the temperature sensitive CH$_4$ and H$_2$O absorption
  bands in $J$ and $H$ bands, which are significantly weaker in models at
  temperature higher than 800\,K than they are in the observed
  spectrum. Conversely, they are too strong in models at a temperature cooler
  than 650\,K.
\end{itemize}

\begin{table*}
\caption{Photometry and  spectrophotometry of CFBDSIR2149 (using
  WIRCam/MegaCam filter set to generate synthetic colours). $z'$ is in
  AB system, all other in Vega system. Spectrophotometry is anchored on
  $J$=19.48$\pm$0.04 from WIRCam photometric measurements. Calibration
  factors are the factors to apply to the spectra in $Y, J, H$ and $K$
  bands so that it matches the
  broad band photometry. $^*$Strong
  systematical uncertainties because of data rescaling below 1$~\mu m$. \label{spec_phot}}
\begin{tabular}{c|c|c|c|c|c|c|c} \hline \hline 
    & $z'_{ab}$  & $Y$ &  $J$ &  $H $ &  $K_s $& $CH_{4off}$ & $CH_{4on}$  \\ \hline
Photometry  & $>23.2$ & 20.83$\pm$0.09 & 19.48$\pm$0.04 &  19.89$\pm$0.11 &
19.35$\pm$0.09 & - &20.7$\pm$0.25 \\ \hline 
Spectrophotometry& 24.51$^*$ & 20.89$^*$ & Reference &  19.98 & 19.43 & 19.35  & 20.89 \\ \hline 
Calibration factor& - & -10\%$^*$ & Reference & +8\%& +8\%  &  - & - \\ \hline
\end{tabular}
\end{table*}

\section{Kinematic analysis: does CFBDSIR2149 belong to the young
  moving group AB Doradus?}
\subsection{Proper motion}
We used the multi-epoch images described in the previous section to
derive the proper motion of CFBDSIR2149. To improve our astrometric
accuracy, we did not use the absolute positions of the source measured
on each image, but calculated a relative local astrometric solution
for each pair of first epoch and second epoch measurements. This was
achieved by cross-matching first and second epochs and calculating the
astrometric solution \citep[using Scamp][]{Bertin.2006} of the second epoch
using the first epoch image as the reference. \\
The most significant source of error was the centroid 
positioning error of this faint source, thus we used PSF-fitting with
\textit{Sextractor} to improve our accuracy. Given the very
asymmetrical spectral energy distribution of late T dwarfs another
source of error is the Atmospheric
Chromatic Refraction \citep[ACR, see e.g. ][]{Dupuy.2012}. Since the
 flux barycenter in broad band filters is not at the same
 wavelength position for T dwarfs than it is for the background stars
 we used to derive the astrometric solution, ACR introduces a
 systematic shift in the centroid position. This depends mostly on the
 filter used and on the airmass at the time of the observations. This effect is
 relatively small in the NIR \citep[a few $mas$,][]{Albert.2011}, but we caution
  our error  
 estimates for proper motion measurements do not take this systematic
 error into account and hence those errors represent lower limits. We
 also neglect parallax effects, which 
 should be below 10$mas$, given the photometric distance (35-50pc) and
 our 28 months baseline between our first 
 (13/08/2009) and our last (23/12/2011) epochs. According to
 \citet{Tinney.2003} the bias from ACR is negligible in $J$-band, and
 therefore, both the parallax effect
 and ACR effects are negligible compared to our measurement accuracy
 ($\sim$30mas.yr$^{-1}$), meaning the uncertainties are dominated by
 measurements errors and not by systematics for our $J$-band
 data. This proper motion 
 measurement are 
 given in Table \ref{pm}, together with the least square linear fit of
  proper motion using as input for each epoch
   the measurement of the position relative to the first epoch. The fit
   was weighted at each epoch by the inverse of the error
   squared. Since we are only interested by the relative motion, we
   assumed the first epoch measurement error to be infinitesimal and affected
   the full error of each $epoch~n~-~epoch~1$ measurement to epoch $n$.
 The measurements using the
 other filter sets are much more dispersed, but as shown on Table
 \ref{pm}, including all data in the fit decreases the fit measurement
 error, though the use of different filters should add small ACR systematics. 
In the following, we use the proper motion derived from the full
data set, but caution that the error bar associated is a lower
limit. A-contrario, the error we derived on $J$-band data only, with
well controlled systematics and much less data, can be
seen as a conservative upper limit of the error bars of the full
data set.

\begin{table*}
\caption{Proper motion measurement of CFBDSIR2149. Discovery position
  (13/08/09) is RA: 21h49$\arcmin$47.2$\arcsec$ Dec: -04d03m08.9s. \label{pm}}
\begin{tabular}{c|c|c|c} \hline \hline 
Epoch 1 & Epoch 2 & Proper motion(RA) & Proper motion(Dec) \\
        &        &    ".yr$^{-1}$    &    ".yr$^{-1}$     \\ \hline 
J$_{13/08/09}$ & Js$_{23/09/10}$       & 0.121$\pm$ 0.042     &   -0.106$\pm$ 0.044 \\ \hline 
Js$_{23/09/10}$ & J$_{26/12/11}$   &   0.026$\pm$ 0.070      &   -0.098$\pm$ 0.088 \\ \hline 
  J$_{13/08/09}$ & J$_{26/12/11}$     & 0.067$\pm$ 0.030     & -0.104$\pm$ 0.043 \\ \hline 
\multicolumn{2}{l|}{Weighted fit (J-band data)} &  0.085 $\pm$ 0.024  &     -0.105   $\pm$  0.031 \\ \hline 
\multicolumn{2}{l|}{Weighted fit (all data)}    &  0.081 $\pm$  0.017   &-0.124  $\pm$  0.019  \\ \hline 
\end{tabular}
\end{table*}

\subsection{Young moving group membership  probability }
Since the photometry and the spectrum of CFBDSIR2149 show tentative
youth indicators, we have estimated the probabilities that this object
is a member of several moving groups and associations
($\beta$-Pictoris, Tucana-Horologium, AB Doradus, TW Hydrae
\citep{Zuckerman.2004} ,Columba, Carina, Argus, the Pleiades,
$ \epsilon$ Chameleontis, the Hyades and Ursa Majoris
\citep{Torres.2008}), as well as a field member. 

\begin{figure}
 \caption{ 
   Vectors representing the norm and direction of the
   proper motion of CFBDSIR2149 compared to the map of ABDMG members
   proper motions.
 \label{PMmaps}}
\begin{tabular}{c c}
\includegraphics[width=9cm]{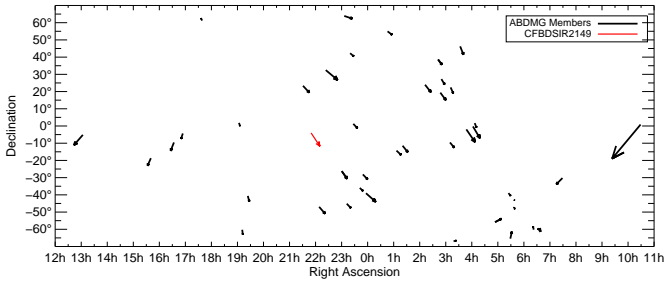}
\end{tabular}
\end{figure}

To do this, we used a Bayesian inference method which consists in
answering to the following question : given the XYZ galactic position and
UVW space velocity of our object  as well as the distribution of
equivalent quantities for each group and the field, what is the
probability it is actually a member of each group ? The field and
groups correspond to the different hypotheses and in principle, the
input parameters to such an analysis would be the XYZUVW for our
object. However, we do not have a measurement for its radial velocity
nor its trigonometric distance. There is a standard way to deal with
this which is generally called {\it marginalisation}
over unknown parameters, consisting of repeating the analysis with
several values for this parameter, then comparing the sum of the
resulting probability densities for each hypothesis. For more details,
we refer the reader to \citet{Feigelson.2012} for a general description of
Bayesian inference, as well as Malo et al. (submitted) for an
analysis similar in many ways to what we do here. We used the  
same XYZUVW distributions for each group and the field as described in Malo et al. (submitted).  

We will highlight the three major differences between this analysis
and ours: (1) we took into account the
measurement errors on input parameters, by convolving each
association's parameter distribution with a Gaussian of characteristic
width corresponding to this error. (2) We treated distance and radial velocity as marginalized
parameters, instead of just distance. (3) We took into account what is
called the prior probability, which was set to unity in the referred
work. The value of this prior
corresponds to the probability our 
object is a member of a given group if we have absolutely no input
data on the object. This is the simple ratio of the
number of members to the group in question with the total number of
stars we could have observed, reflecting the obvious fact that 
any random star had a much higher probability to belong to the field
than to any young association, since there are many more field stars
that there are young associations members. 
If this quantity is accurately estimated, it would ensure that 90\% of
members with 90\% membership probability would be actual members. 
 Without this prior,
CFBDSIR2149 would have a 99.9\% membership probability to belong 
either the AB
Doradus moving group (ABDMG) or to the $\beta$-Pictoris moving group
(BPMG). Notably, the membership probability to ABDMG is very high
because of the proximity of CFBDSIR to ABDMG cinematic locus (see
Figure \ref{PMmaps}).\\ 

To give a value to this prior for the young associations hypotheses, we used the number of known young
moving group members.  For the
field hypothesis, we used the fact our object is conservatively younger than 
500 Myr since its spectrum is rather indicative of an
age in the 20 to 200\,Myr range. Given that
evolutionary and atmosphere models of young substellar objects are not
yet fully reliable, we preferred to set a conservative upper age
limit of 500\,Myr. To inject this information in our Bayesian
inference, we simply 
treated the prior as the ratio of stars in each moving group with the
total population of field stars younger than 500 Myr distant from 0 to
100 pc in a galactic disk simulation from Besan\c{c}on Galaxy
model \citep{Robin.2003}. We are confident this
range of distances is reliable, since the young moving groups we
considered lie within 100 pc of the Sun and the photometric distance
of our candidate is comfortably below 100 pc. This estimate is in fact
quite conservative since it is generally accepted that the populations of 
young moving groups are still incomplete, meaning we might
underestimate their respective membership probabilities. A third
reason makes our membership probability conservative: when
considering the field hypothesis we don't use the
proper motion distribution of specifically young field objects, but of
the general field population. Since young field objects have a
narrower proper motion distribution \citep{Robin.2003}, we therefore
overestimate the field membership probability.\\

With this Bayesian analysis that conservatively takes into account
that young field 
stars are much more numerous than young association stars, we find
CFBDSIR2149 has a membership probability of 79.4\% for AB Doradus and 13.3\%
for the field. The third most likely
hypothesis, a membership to the young moving group $\beta$Pictoris,
has a 7.3\% probability.

\begin{figure}
 \caption{Bayesian probability densities over distance and radial
   velocity derived from CFBDSIR2149 proper motion, for the 3 most
   probable membership hypotheses. The contours encircle respectively
   10\%, 50\% and 90\% of the  probability density for each hypothesis.\label{bayes}}
\includegraphics[width=9cm]{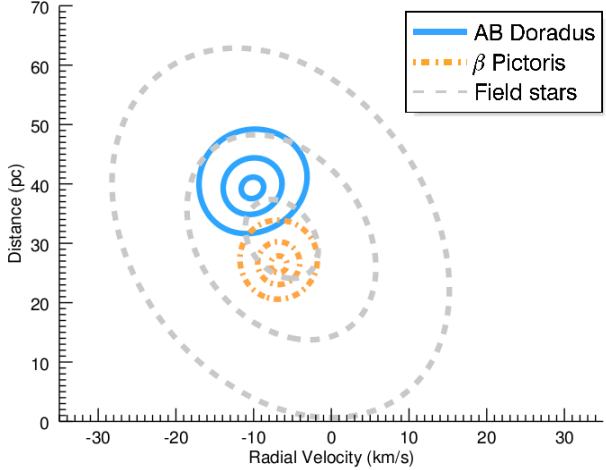}
\end{figure}

This makes CFBDSIR2149 a good candidate member to the 50-120
\,Myr old, solar metallicity, ABDMG moving group
\citep{Zuckerman.2004,Luhman.2005,Ortega.2007}. The most likely values of the marginalized parameters (distance and
radial velocity)  for the most probable hypotheses (AB Doradus,
field, and Beta Pictoris membership) are shown on Figure \ref{bayes}.
For the young field hypothesis, the Bayesian estimate
provide a statistical (most likely) distance of $31\pm13$ pc and a radial
velocity of $-6 \pm 9$ km.s$^{-1}$. According to the
\citet{Allard.2012} BT-Settl isochrones, the photometric distance of a 700\,K,
500\,Myr old field brown dwarf with CFBDSIR brightness would be between 25
and 40\,pc (respectively using the $K_s$ and $J$-band photometry), in
reasonable agreement with the Bayesian kinematic estimation. A younger
age -our field prior is compatible with any age below 500Myrs- would lead to higher photometric distances and therefore
to a more marginal agreement for the field hypothesis.

For the highest probability hypothesis, namely that our target
belongs to ABDMG, the statistical distance  is $40\pm4$ pc
and radial velocity of $-10 \pm 3$ km.s$^{-1}$. According to the \citet{Allard.2012} BT-Settl
isochrones, the photometric distance of a 700\,K,
120\,Myr old brown dwarf with CFBDSIR brightness would be between 35
and 50\,pc (respectively using the $K_s$ and $J$-band photometry), which
is in good agreement with the Bayesian estimation.

 For the BPMG hypothesis the Bayesian estimate
provide a statistical (most likely) distance of $27\pm3$ pc and a radial
velocity of $-7 \pm 2$ km.s$^{-1}$. According to the
\citet{Allard.2012} BT-Settl isochrones, the photometric distance of a 650\,K,
20\,Myr old field brown dwarf with CFBDSIR brightness would be between 25
and 35\,pc (respectively using the $K_s$ and $J$-band photometry), in
reasonable agreement with the Bayesian kinematic estimation. 

If we integrate the probability distribution of Fig.\ref{bayes} over
the photometric distance estimate range, we can obtain a membership
probability that include this information. We use the
  photometric distance estimate for the field for an age of 500Myr
  which is the closest to the field Bayesian estimate, and therefore
  slightly favours the field hypothesis.  This yields a
  membership probability of 87\% for the ABDMG, 7\% for the BPMG, and
  6\% for the field. Also, since a fraction of field objects are younger than
  150Myr, the actual probability that CFBDSIR2149 is actually younger than
  150Myr is higher than its membership probability to ABDMG and BPMG.  Taking that
  into account, \emph{CFBDSIR2149 has a
  membership probability of 87\% for the ABDMG, 7\% for the BPMG, 3\%
  for the young field ($age<$150Myr) and 3\% for the field (150Myr$<age<$500Myr).}

We caution that this
introduces significant uncertainties into the calculation, because of
the low reliability of photometric distance estimates.
 To dampen this
concern, we derive in the following what could be seen as lower limits of ABDMG
membership probability by looking at scenarios even more favourable
to the field, using a prior allowing
  objects as old as 1Gyr and 10Gyr  to be included in the prior. This is very
  conservative because CFBDSIR2149 spectrum indicate a much younger
  age ($<$200Myr, see previous section).  With
  the prior including all field stars younger than 1Gyr, this yields 72\% membership
  probability to ABDMG, 22\% to the field of all ages below 1Gyr, and
  6\% to the BPMG, based on position and
  proper motion alone. If we integrate the
  Bayesian distance estimates on the photometric distance estimate
  ranges, we derive a 
  membership probability of 83\% for the ABDMG, 6\% for the BPMG, 3\%
  for the young field ($age<$150Myr) and 8\% for the field (150My$<age<$1Gyr).
  If we use the even more extreme prior including all field stars
    younger than 10Gyr and integrate these probabilities on the
    photometric distance estimates (using a distance range of 20-35pc
    for a 5Gyr T7 dwarf), we still get 63\% membership
  probability to ABDMG, 32\% to the field of all ages below 10 Gyr, and
  5\% to the BPMG.  As discussed in the previous section,
  the spectral features of this object are not compatible with an old
  age, unless it has a strongly super-solar metallicity. Since high
  metallicity stars are rare, and are especially rare among old stars,
  using as weight of our field hypothesis all stars below 10Gyr is not
  a realistic hypothesis. Using this prior does therefore provide a
  very strong lower limit on the membership probability of CFBDSIR2149
  to the ABDMG, but is not indicative of its actual membership
  probability. 
 
Our Bayesian analysis taking
into account right ascension, declination, proper motions and distance
estimates allow to set the following constraints on the membership
probability of CFBDSIR2149 to the ABDMG:
\begin{itemize}
\item An extremely conservative lower limit of 63\% that we do not
  consider realistic. 
\item A very conservative lower limit of 83\%. 
\item A conservative 87\% membership probability of
  CFBDSIR2149 to the ABDMG, that we will use as our best estimate in
  the following.
\end{itemize}

We note that when including the possibility that
  CFBDSIR2149 is a field object younger than 150Myr but unrelated to a known
  association, the overall probability that it is  younger than 150Myr
  is therefore higher than 95\%. The following section explores the
  spectral properties of CFBDSIR2149 assuming that it is indeed
  younger than 150Myr. 

  Finally, as a sanity-check, we determined
  whether another CFBDS brown 
  dwarf, CFBDSIR1458, described in \citet{Delorme.2010} as an old
  brown dwarf on the ground of its spectral features, could turn out
  as a young moving group member using the same kinematic
  analysis. The result is a 100\% probability that this object belongs to the
  field, with the probability to belong to any young moving group lower
  than $10^{-20}$, even using the 500Myr prior, which is more
  favourable to young moving group membership. These kinematic results
  are therefore in
  excellent agreement with the spectral analysis by
  \citet{Delorme.2010,Liu.2011} identifying CFBDSIR1458 as an old field
  brown dwarf.

\section{Physical properties of CFBDSIR2149}
\subsection{Comparison to other late T dwarfs}
%

\begin{figure}
 \caption{Comparison of BT-Settl 2012 models at the same temperature
   (700K), but with different gravity (red:log g=3.5, black: log
   g=4.5). The KI doublet position is highlighted by the dashed
   lines. \label{Kdoub_theo}}
\includegraphics[width=6cm,angle=270.]{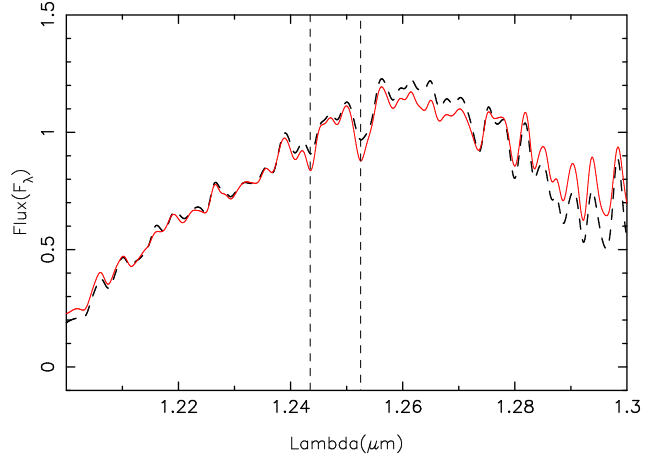}
\end{figure}

 A direct comparison of CFBDSIR2149's spectrum with the spectra of other T7-T8.5 brown dwarfs (see
  Fig. \ref{spectra}), clearly shows its enhanced $K$-band flux. The
  spectrum of CFBDSIR2149 shows another low 
 gravity feature, namely the enhanced absorption by the 1.25$~\mu m$
 potassium doublet (see Fig. \ref{Kdoub_theo}). According to
 \citet{Allard.2012sub}, the strength of the doublet in the spectra
 depends both on the abundance of potassium in the atmosphere, and on
 the strength of the CH$_4$ and H$_2$O absorptions bands that
 shape the pseudo-continuum through which the doublet forms. At T
 dwarf temperatures, a
 lower gravity increases the doublet strength by playing on
 each of these parameters: the pseudo-continuum is weakened at low
 gravity for a given effective temperature while the potassium
 abundances tends to increase because of enhanced vertical mixing effects in
 low gravity atmospheres. The same trend has been empirically observed
 in T dwarf by  \citet{Knapp.2004}. 

\begin{figure*}
 \caption{Comparison of CFBDSIR2149 spectrum (red) in the $J$ band, with low gravity
   \citep[bottom, ][]{Leggett.2002,McLean.2003} and regular  \citep[top,
   ][]{Strauss.1999,McLean.2003} T6(left) and T7 (right) brown
   dwarfs (black). The dashed line indicate the position of the potassium
   doublet. \label{Kdoub}}
\begin{tabular}{c c} 
\includegraphics[width=8cm,angle=0.]{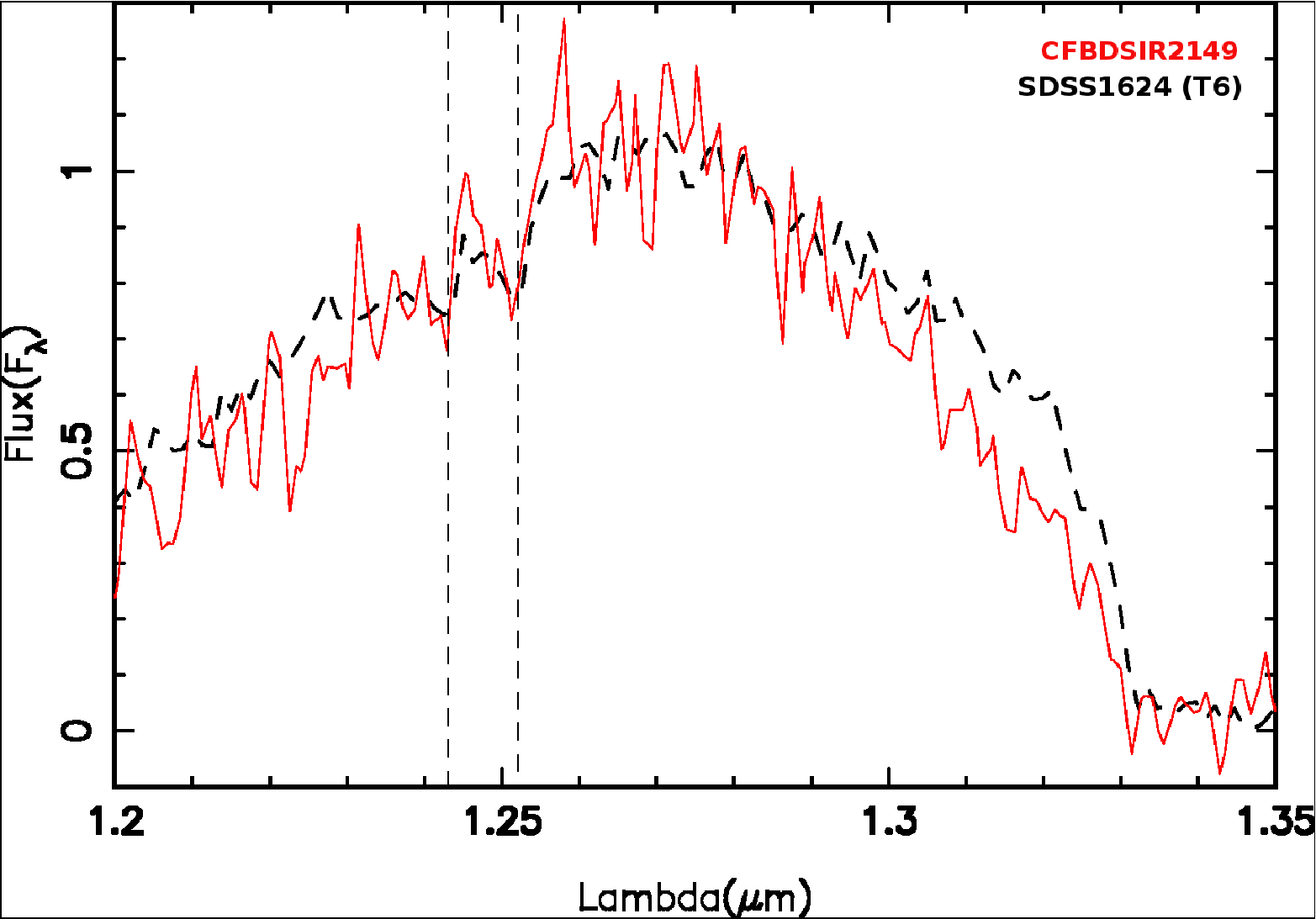}
&
\includegraphics[width=8cm,angle=0.]{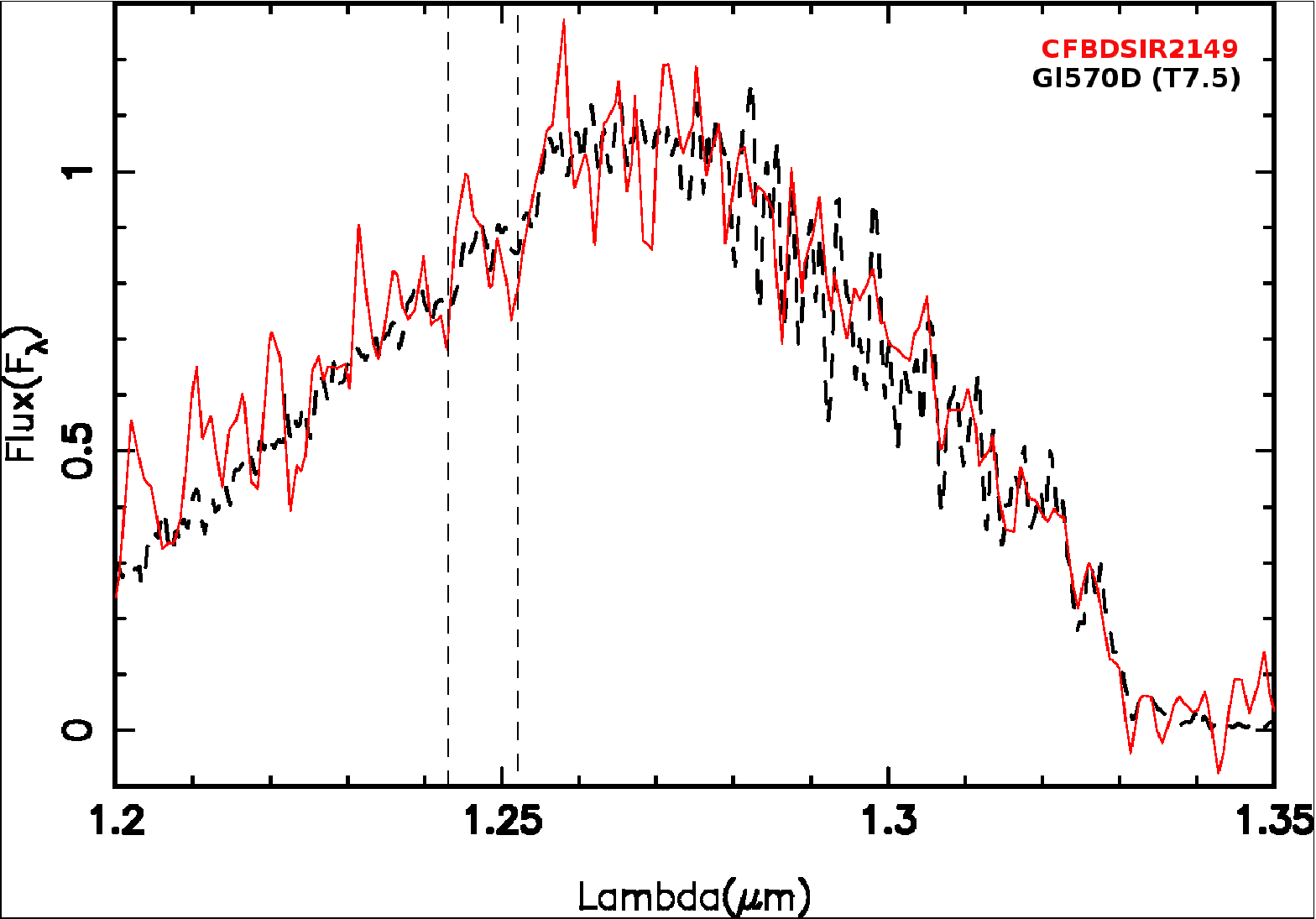} \\
\includegraphics[width=8cm,angle=0.]{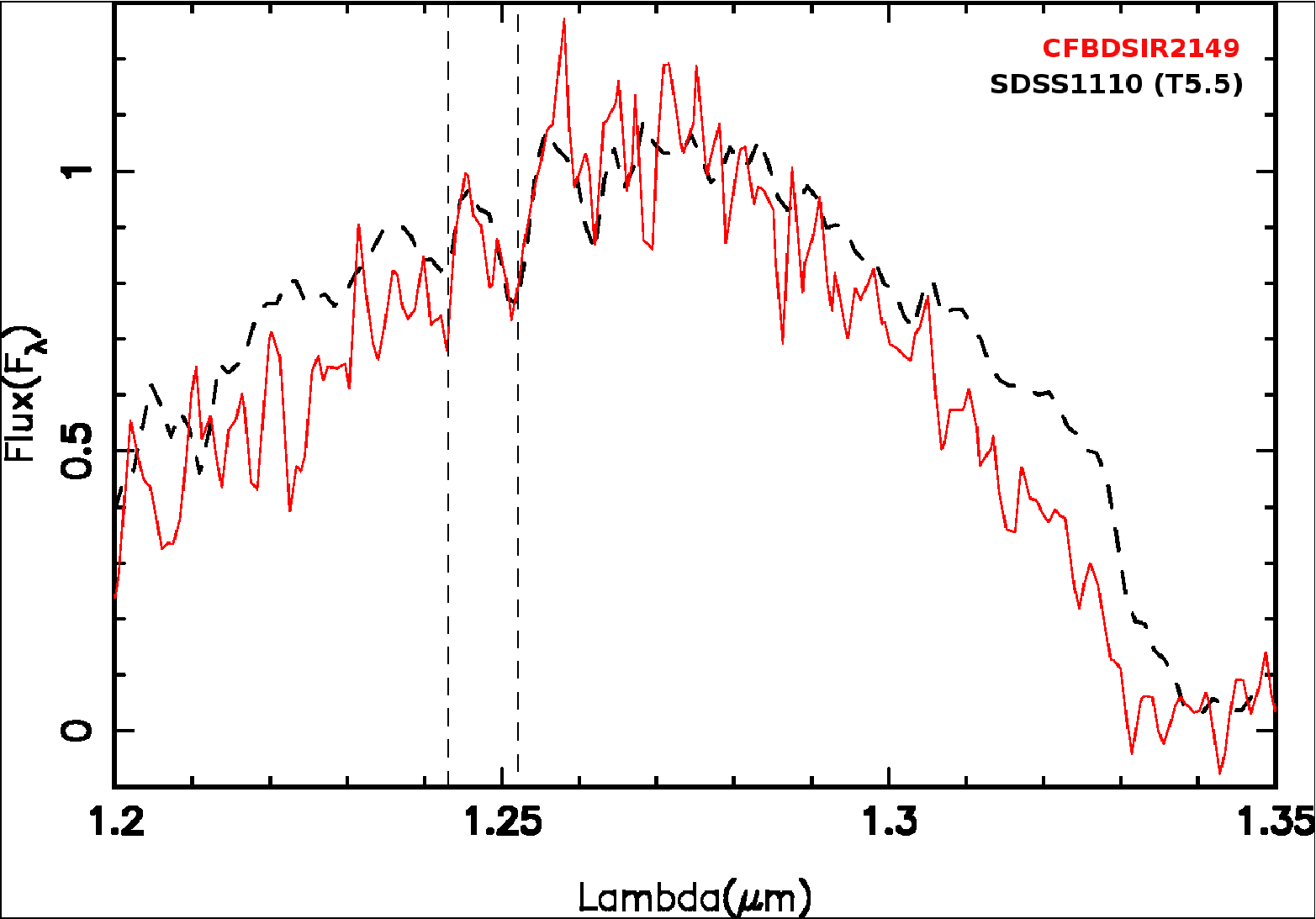}& \includegraphics[width=8cm,angle=0.]{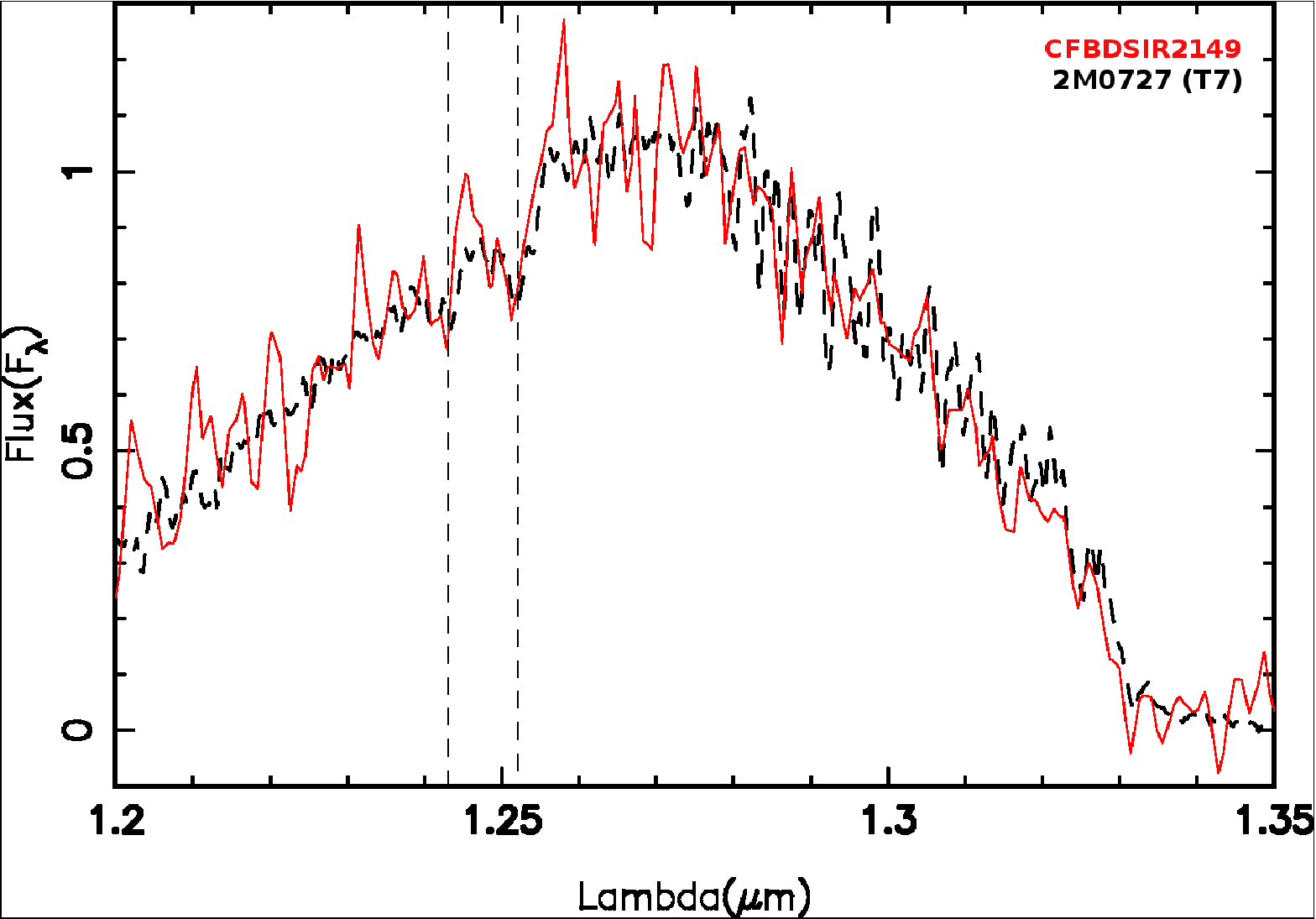}
\end{tabular}
\end{figure*}

\begin{figure*}
 \caption{Comparison of CFBDSIR2149 spectrum with other Late T
   dwarfs. Spectrum have been taken from \citet{McLean.2003,Burgasser.2003,Chiu.2006,Burningham.2011a} \label{spectra}}
\begin{tabular}{c c} 
  \includegraphics[width=8cm,angle=0.]{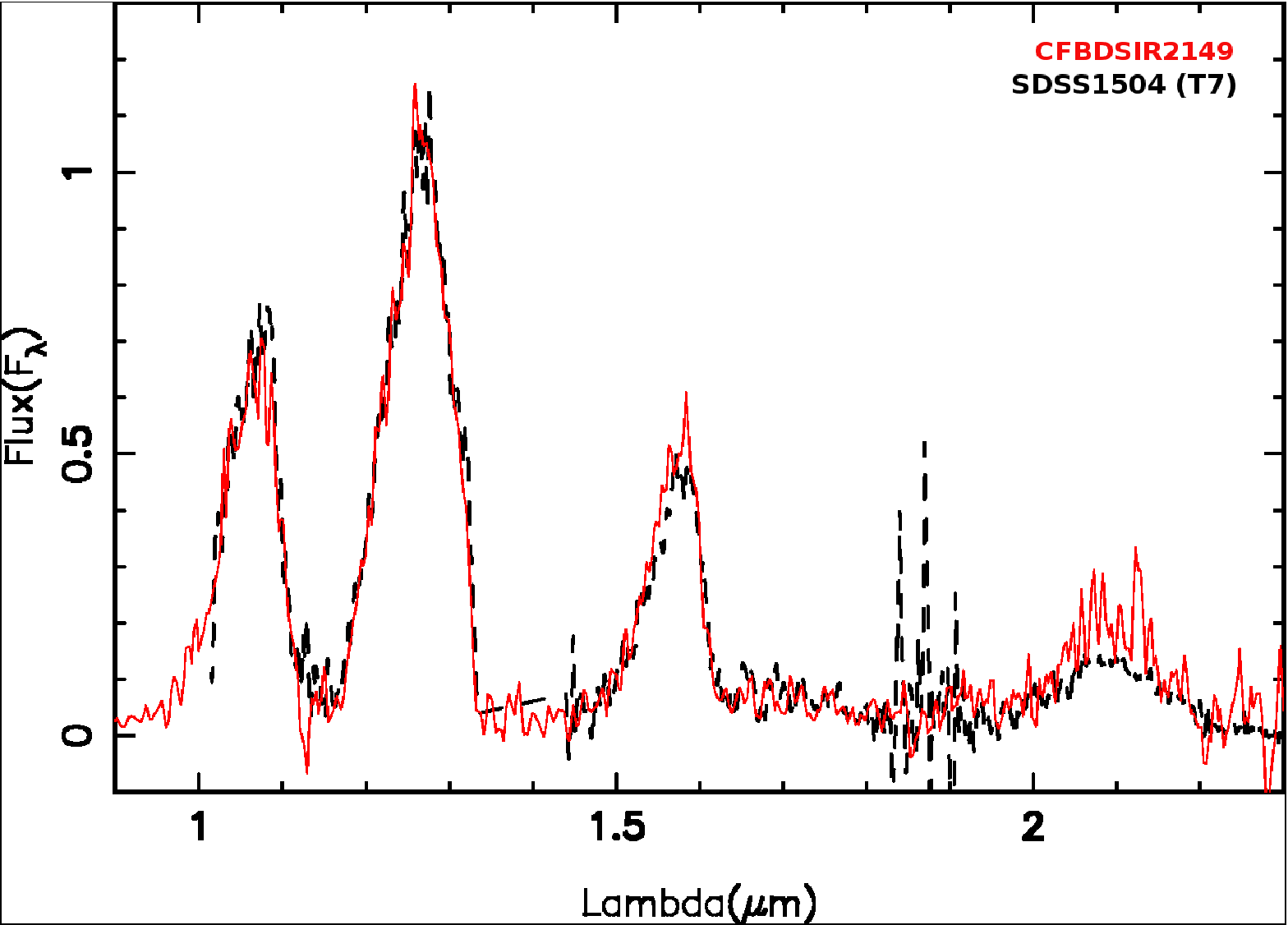}  &
 \includegraphics[width=8cm,angle=0.]{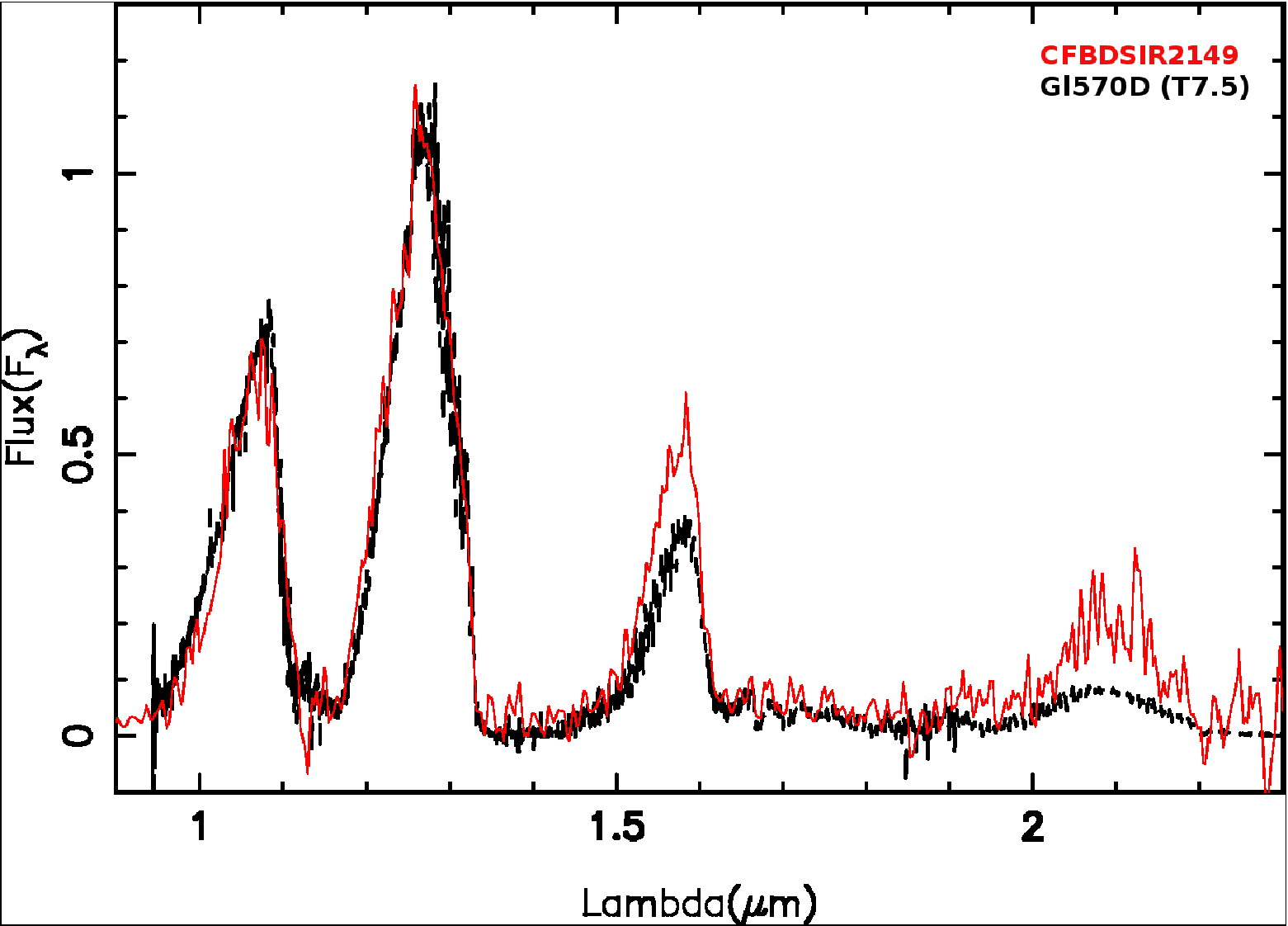} \\ 
  \includegraphics[width=8cm,angle=0.]{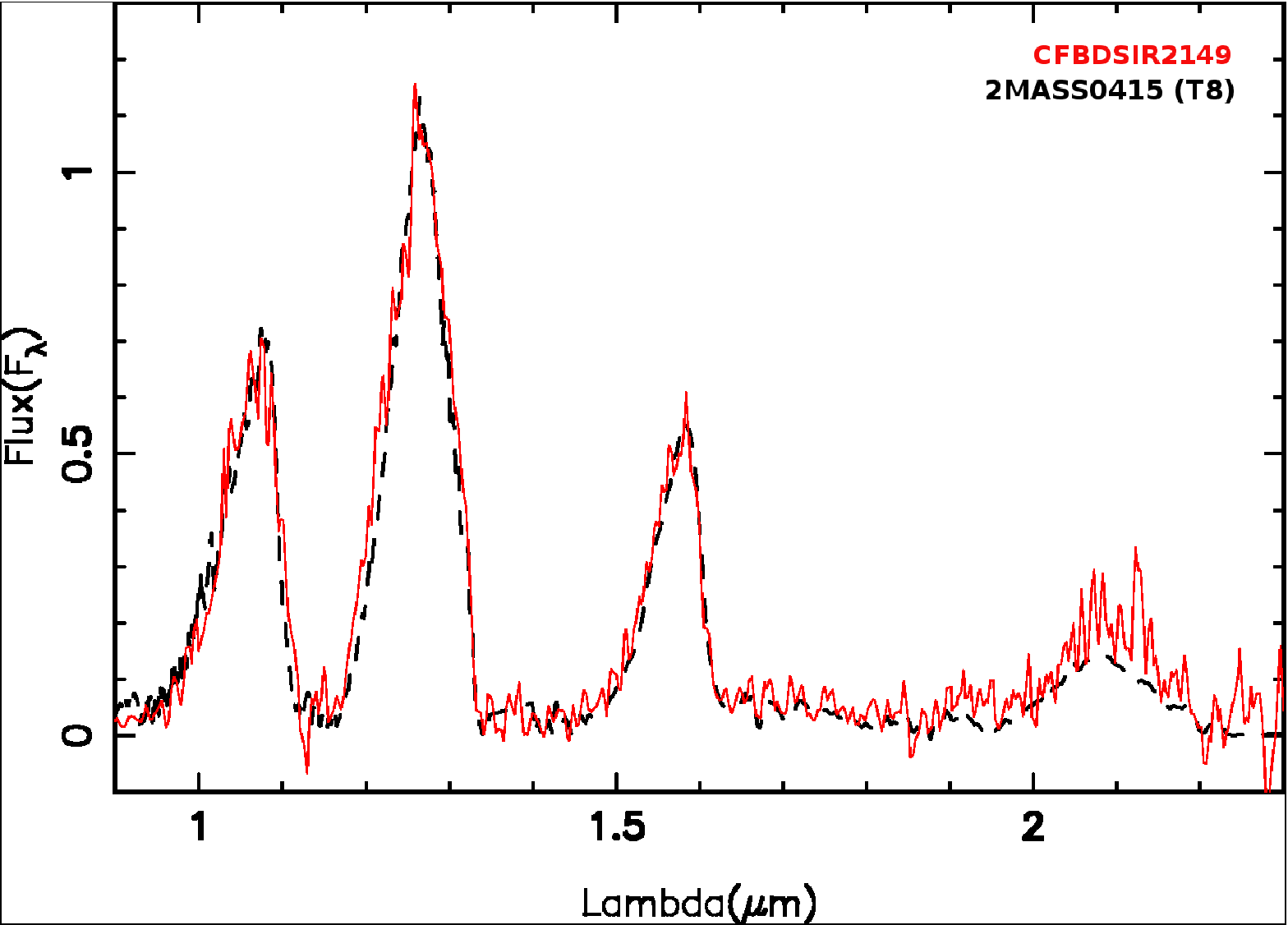}  &
  \includegraphics[width=8cm,angle=0.]{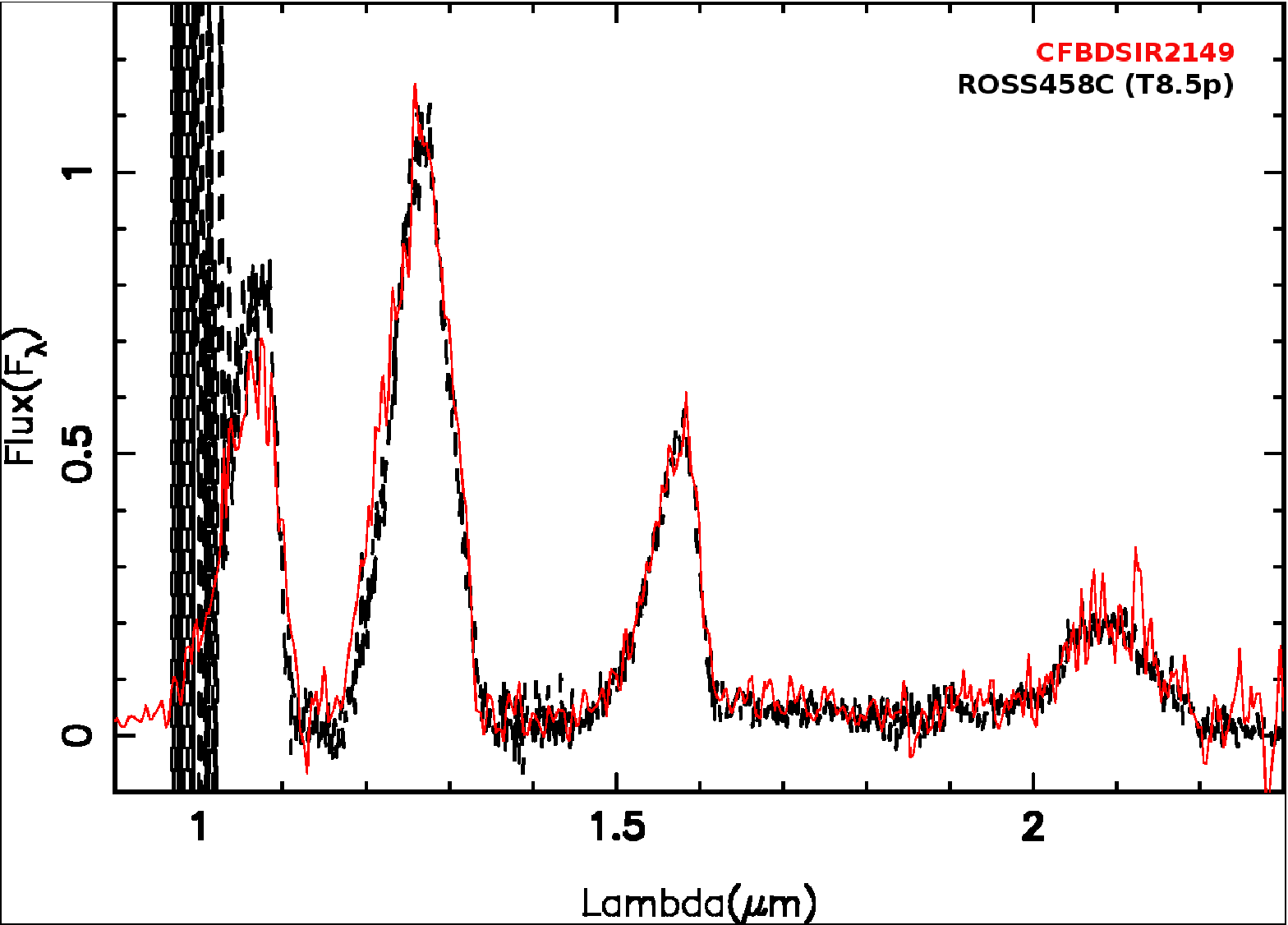} \\
 \end{tabular}
\end{figure*}

This doublet decreases in strength from T4 to T7, until it totally disappears on
regular field dwarfs later than 
 T7 \citep{Knapp.2004}, and should be almost absent of the T7/T7.5
 spectrum of CFBDSIR2149 if it 
 had field gravity and solar metallicity. The comparison with the T7
 2M0727+1710 (See Fig. \ref{Kdoub}), which has already been identified as a low gravity T7 by
 \citet{Knapp.2004} illustrates that 
 the KI absorption doublet is more prominent in CFBDSIR2149. \\

Another interesting object to compare with is the very red T8.5 brown dwarf
  Ross458C
  \citep{Scholz.2010,Goldman.2010,Burgasser.2010,Burningham.2011b}
  which presents a $K$-band flux excess comparable to what we observe for CFBDSIR2149. As
  discussed  in \citet{Burgasser.2010,Burningham.2011b}, this object
  shows many features of low gravity and belongs to a young stellar system (150-800\,Myr), with a
  slightly higher than solar metallicity ([Fe/H]$\sim$0.2-0.3).

 Before we engage in a closer comparison between CFBDSIR2149 and
 Ross458C, we have to point out that since CFBDSIR2149 appears
 slightly warmer than Ross458C, it should have, because of effective
 temperature alone, a redder $J-K_s$ (See Fig.\ref{colspt}) and a higher
 J/K spectral index. This
 makes the direct comparison of these gravity-sensitive features
 difficult: they are slightly stronger in CFBDSIR2149 but this
 difference, including the presence of the 1.25$~\mu m$ potassium
 doublet, might be fully accounted for by the difference in 
 effective temperature rather than by an even lower gravity.  In order
 to compare the colours of both 
 objects using the same photometric system, we calculated the NIR
 colours of Ross458C in the WIRCam filter system using the spectra
 provided by \citet{Burningham.2011b}. The resulting colours are shown
 in Table \ref{2149_458} and are very similar to those observed
 for CFBDSIR2149. Note that the $J-K$ colour of Ross458C with WFCAM
 filter system is much bluer \citep[$-0.25$, ][]{Burningham.2011a}
 because the atypical $K$-band filter used on WFCAM cuts a
 significant part of T dwarf flux blueward of 2$~\mu m$ and extends
 red-ward of 2.3$~\mu m$ where late T dwarfs have no 
 flux left.\\

   Because of the degeneracy between effects of low gravity and high
  metallicity, the similar colours and spectral features of
  CFBDSIR2149 with respect to Ross458C can be explained either by a
  similar age/[M/H] combination or by a younger age and a more common
  solar metallicity for CFBDSIR2149. There is also 
  the possibility that CFBDSIR2149 is older and has an even higher
  metallicity than Ross458C but this is much less probable both
  because of its 
  kinematics and because of the scarcity of significantly
  over-metallic stellar systems \citep[only $\sim$ 4\% of stars have
    metallicity$>$0.3, ][]{Santos.2005}.  Given the probable association of
  CFBDSIR2149 to the 50-120\,Myr AB Doradus moving group,
  the hypothesis that it is a slightly warmer, slightly
  younger solar metallicity counterpart of Ross458C is most
  likely, especially because it is also consistent with the
  [M/H]=-0.02$\pm$0.02 metallicity distribution derived by
  \citet{Ortega.2007} for ABDMG members. According to the Lyon
  stellar evolution models 
  \citet{Baraffe.2003}, at solar metallicity, such a 650-750\,K brown dwarf
  aged 50\,Myr has a mass of 4 to 5 Jupiter masses and a log\,g of about 3.9. This would make
  CFBDSIR2149 a ''free-floating planet", with the
  same atmospheric properties as any 50\,Myr old, 4-5\,M$_{Jup}$,
  T-type exoplanet, and therefore an invaluable benchmark for
  exoplanets atmospheres studies.  In the case we would adopt the higher age
  hypothesis of 120\,Myr \citep{Luhman.2005,Ortega.2007}, the
  conclusion would be similar, CFBDSIR2149 then being a 120\,Myr old,
  6-7\,M$_{Jup}$ free-floating planet with a log\,g of $\sim$4.1.\\
 
\subsubsection{Revisiting Ross458C}
The young age hypothesis for Ross458C is supported by the fast
 rotation \citep[11km.s$^{-1}$; ][]{Donati.2008} observed
 for its primary M1 star Ross458A. The corresponding upper limit of the period (in case the projected rotational speed
 $V.sin(i)= V$, the equatorial rotation speed) of
 $\sim$2.5\,days. \citet{Delorme.2011} established that early M dwarfs in the
 Hyades and Praesepe had already converged toward a clean mass-period
 relation. Gyrochronology \citep{Barnes.2003} therefore allows
 asserting that almost all early M dwarfs which rotate faster than the 12-14
 days observed in the early M dwarfs of these $\sim$600\,Myr clusters
 are younger than the 
cluster themselves. The exceptions are the few stars which experienced
 significant angular momentum transfer from orbital momentum to
 rotational momentum through close tidal interactions within a
 multiple system. Since Ross458A has a rotation period much shorter than 12-14
days, and is not known as a very close dynamically interacting binary, we
can safely deduce that the Ross458 system is younger than
600\,Myr. This upper age limit is more stringent than the
 800\,Myr upper age limit from
\citet{Burgasser.2010} and is consistent with the 200-300\,Myr 
 derived from the spectral analysis of Ross458C and CFBDSIR2149. 

Given the probable higher metallicity and lower temperature of Ross458C, the
gravity of both objects should be similar.  Because
of the fast 
  cooling down of a substellar object during its first few hundreds\,Myr,
  a gravity within 0.2 dex. of CFBDSIR2149, for the
  650-700\,K Ross458C, is only compatible 
  with an age lower than 200\,Myr (300\,Myr if we use the older age
  estimate for the AB Doradus moving group), with a corresponding
  maximum mass of 
  8 (respectively 10) M$_{Jup}$. These rough estimates, also
  compatible with the  age value from \citet{Burgasser.2010},
  would redefine the T8.5 brown 
  dwarf Ross458C as an exoplanet orbiting the Ross458 stellar system,
  if we follow the controverted 
  International Astronomical Union official definition of a planet,
  even if its formation mechanisms are likely more akin to
  the formation of multiple systems. Indeed, the formation of Ross458
  through planetary formation scenarii such as core accretion 
  or gravitational instability is unlikely
  both because of 
  the very large separation \citep[1200 AU,][]{Scholz.2010}, and low mass
  of the M1 primary Ross458A. \\ 


\subsection{Comparison to models}

\begin{figure*}
 \caption{Comparison of CFBDSIR2149 full spectrum at R=225 (red) with
   various models (black). {\bf Upper left:} Burrows Sudarsky Lunine
   2003 model for 
   a 100\,Myr, 7 M$_{jup}$ planet (i.e T$_{eff} \sim$750\,K, log\,g$\sim$4.1) with no clouds. {\bf Upper right:} BT-Settl
   models for Teff=650\,K and log g =3.75. {\bf Lower Left:} \citet{Madhusudhan.2011} cloudy ''E" model at Teff=700\,K, log =4.0 and 60$~\mu m$
   enstatite particles in clouds. {\bf Lower Right:}
   \citet{Madhusudhan.2011} cloudy ''A" model at Teff=700\,K, log =4.0 and 100$~\mu m$
   particles in clouds  \label{Hexcess}}
\begin{tabular}{c|c} 
\includegraphics[width=8cm,angle=0.]{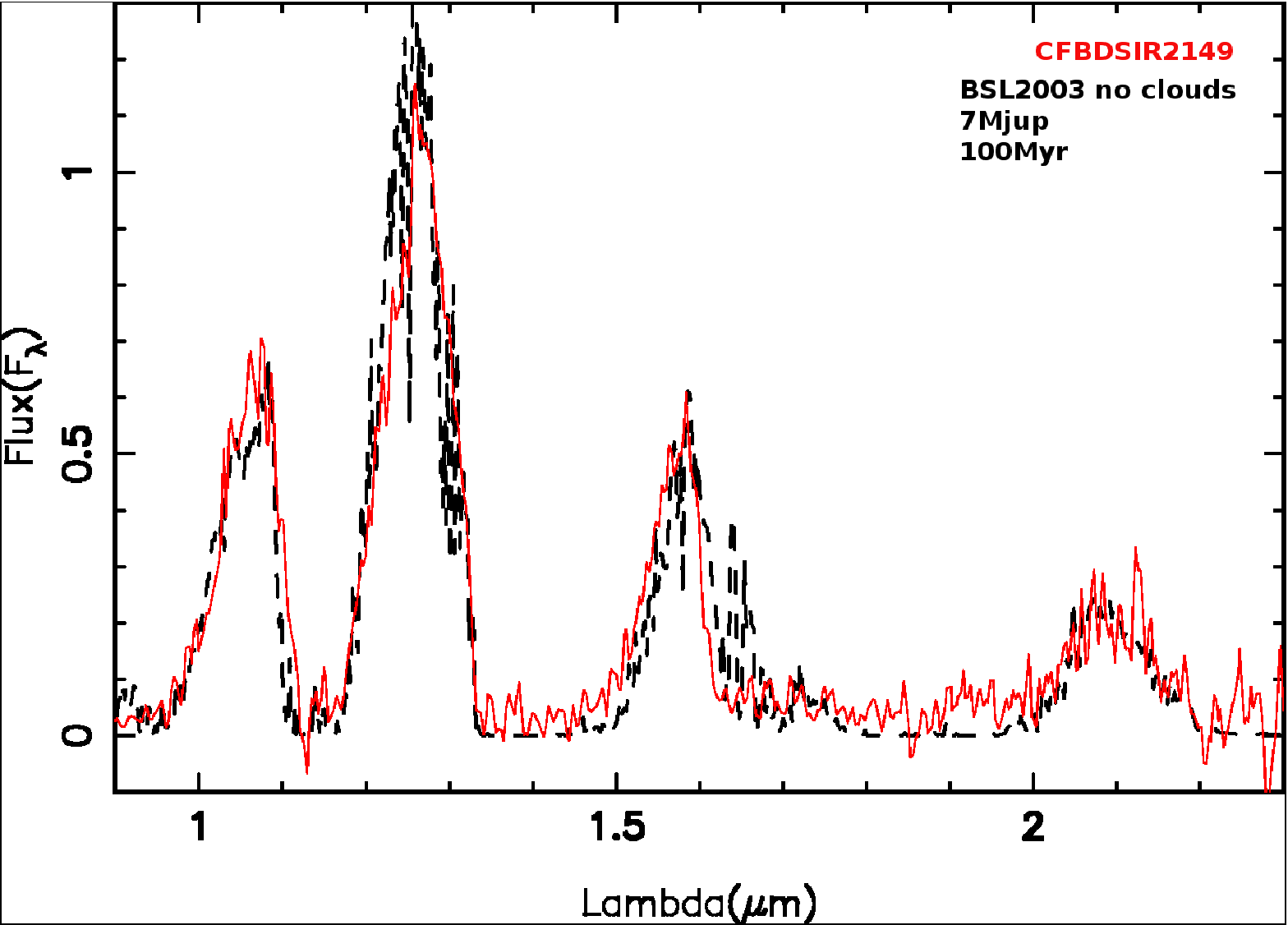} &\includegraphics[width=8cm,angle=0.]{CFBDSIR2149_Vs_lte0065-3_75.ps} \\ 
\includegraphics[width=8cm,angle=0.]{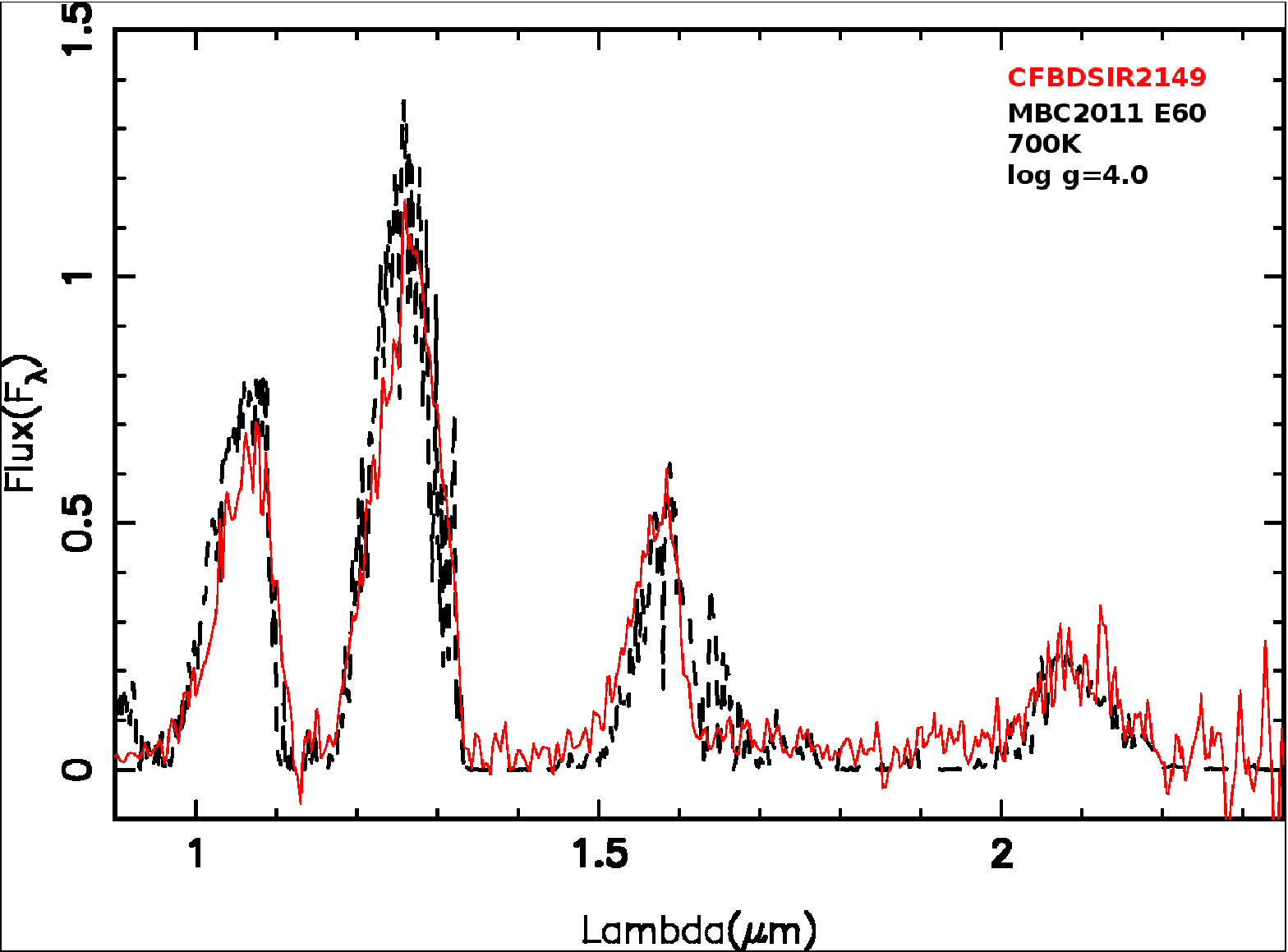} & \includegraphics[width=8cm,angle=0.]{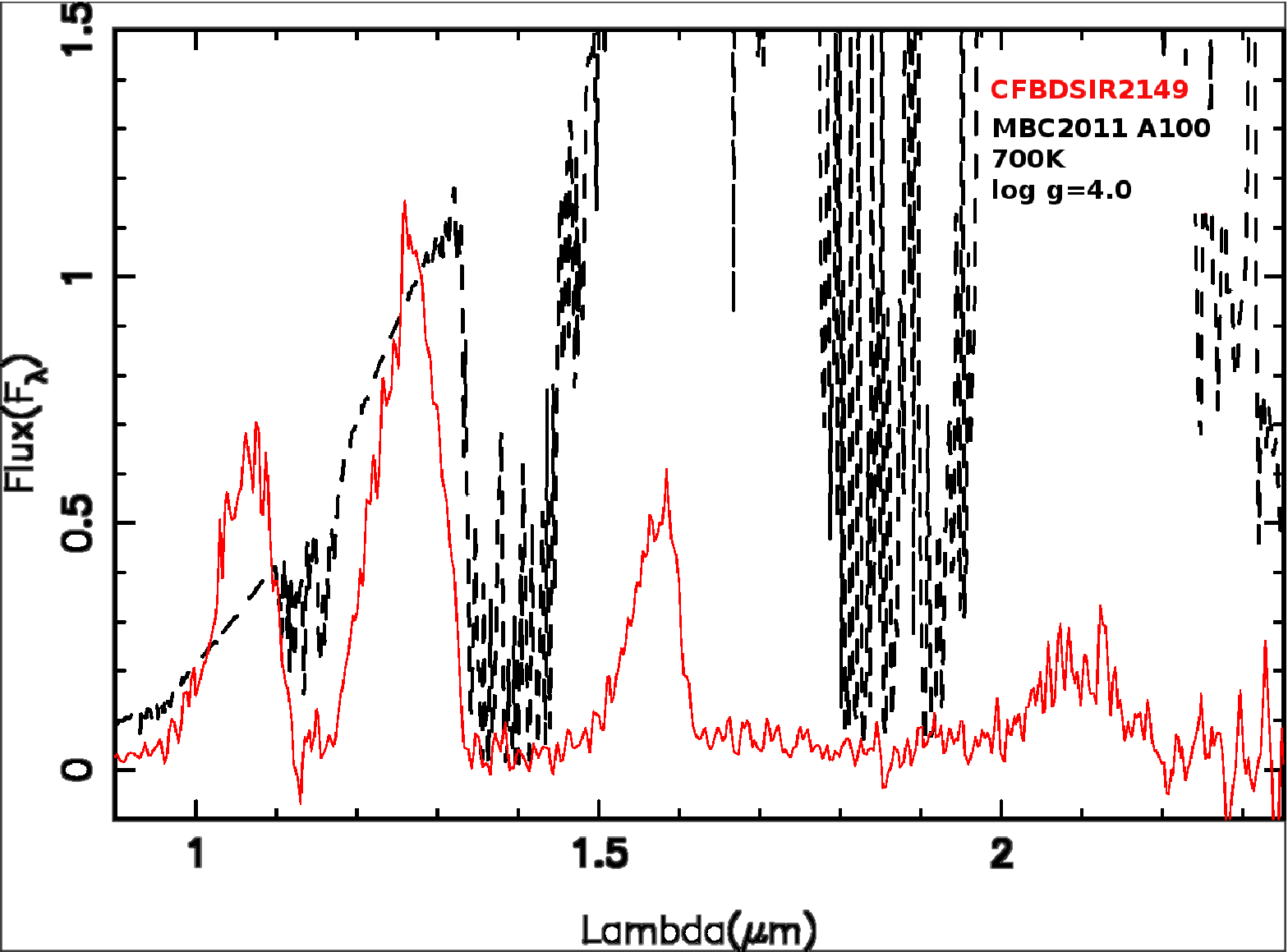}\\
\end{tabular}
\end{figure*}

  We used several sets of models to match the observed spectra of
  CFBDSIR2149. The corresponding plots are shown in
  Fig. \ref{spectraVsmodelsTeff} (BT-Settl models) and
  Fig. \ref{Hexcess}, and include planetary
  atmosphere models meant for the HR8799 planets \citet{Currie.2011} adapted from
  \citet{burrows.2003} (hereafter referred as BSL03) and from
  \citet{Madhusudhan.2011}, 
  (hereafter referred as MBC20111) as well as the
  BT-Settl-models.\\

  We did not attempt a numerical fit of our observed spectra to these
  various models grids because of the strong (and model dependent)
  systematics uncertainties in atmosphere models, notably missing
  CH$_4$ absorption lines 
  blue-ward of 1.6$~\mu m$ whose lack is obvious in
  Fig.\ref{Hexcess}, including right on the $H$-band peak. We preferred
  matching models visually, rather 
  than injecting ad-hoc restrictions in a fit by artificially
  weighting out parts of the models we suppose
  are affected by systematics. Discussing the resulting numerical
  values of such a fit would not be more relevant than discussing the
  models that visually match the observed spectrum and would
  unfairly hide some of the limitations that come with comparing poorly
  constrained models and poorly sampled model grids to noisy data.\\

  The main difficulty to match CFBDSIR2149 spectrum is to find a model
  that correctly reproduces both its very deep CH$_4$ absorption in the
  $H$-band and its strongly enhanced $K$-band flux. The former is
  easily matched by all models with cool (~700\,K) temperatures and relatively
  high gravities (log g=4.5-5.0; see Fig. \ref{Hexcess}) while the latter can be
  matched by warmer, low gravity models (see Fig. \ref{spectra}), but
  the combination of both is 
  difficult to reproduce. The MBC models with clouds of type ''A", ''AE"
  and ''AEE" very significantly fail to reproduce the SED of 
  CFBDSIR2149 (see the ''A" case on the lower right panel of
  Fig. \ref{Hexcess}), because they predict an extremely red NIR SED,
  due to their very thick cloud layers. In the "A" case,  the cloud
  model includes no upper cloud boundary at all, i.e. no dust
  sedimentation effects of any kind, corresponding to the simplified
  cases of fully cloudy atmospheres like the AMES-Dusty models.  The
  "E" models by comparison implement a steeply decreasing condensate
  density above a certain pressure level given as a model parameter,
  and can thus be tuned to produce clouds comparable to the BT-Settl
  models  or the Marley \& Saumon models for a certain
  $f_\mathrm{sed}$.   As seen on Fig. \ref{Hexcess}
  these ''E" cloud models as well as the  BSL2003 and 
  BT-Settl models offer a decent match to the
  spectrum when using relatively cool temperatures (650-700\,K) to depress the
  H-band flux and relatively low gravities (log\,g=3.75-4.1) to enhance
  the $K$-band flux, but all models suffer from missing CH$_4$
  lines blue-ward of 1.6$~\mu m$. This systematic shortcoming
  acknowledged, comparison with models gives a temperature of
  650-750~K and a log\,g of 3.75 to 4.1 for CFBDSIR2149. BT-Settl
  models favour a cool 650\,K, log g=3.75 value that would be only compatible,
  according to the \citet{Baraffe.2003} substellar evolution models
   with hypothesis that
  our target is a member of the 12 Myr-old $\beta-$Pictoris moving group,
  and with ABDMG membership but only for its youngest age estimate
  \citep[50\,Myr;][]{Zuckerman.2004}. Conversely, the ages derived from
  the MBC2011
  ''E" cloud models and  BSL2003 models (also using
  \citet{Baraffe.2003} models to translate temperature and gravity
  into age and mass) would be fully compatible with
  the probable ($>$95\%) hypothesis that CFBDSIR2149 is younger than
  150 Myr. Even in the extreme hypothesis that this object would be aged
  500\,Myr, its mass (11M$_{Jup}$) would still be below the
  deuterium burning mass of 13M$_{Jup}$ that serves as an artificial
  limit between planets and brown dwarfs. According to this
  mass-driven definition, CFBDSIR could be called an IPMO or
  a free-floating planet. \\

  A major remark we can draw from these comparisons is
  that CFBDSIR2149's spectrum is close to
  the standard BT-Settl and BSL2003 models and can be matched without
  using the thick
  clouds models that have been created to account for the 
  peculiar spectral energy distribution of the young exoplanets
  orbiting HR8799 \citep{Currie.2011}. Given the very high dependence of the
  modelised SED to the type of cloud these MBC2011 models (see the
  completely different SED modelised on the last row of
  Fig. \ref{Hexcess}, for the same 
  gravity and temperature but with a different cloud parameter
  setting), the fact that one of the cloud settings matches the
  observed spectrum at some point of the gravity/temperature grid as
  well the BT-Settl or BSL2003 models do is not 
  surprising. \citet{Burningham.2011b}
  also concluded that such adaptable cloud models were not necessary to
  account for Ross458C's spectrum, which can be fitted with BT-Settl
  models which use a self-consistent cloud model over their full
  parameter range. 
 
If CFBDSIR2149 is indeed a 4-7M$_{Jup}$  
  ABDMG member, this would mean that the similar cool exoplanets of
  late T type targeted by upcoming high contrast imaging instruments
  should have a 
  spectrum closer 
  to models prediction than the heavier/warmer HR8799 planets of early
  T type. This would not be surprising, since the L/T transition
  atmospheres are
  known to be much more difficult to model than late T atmospheres for
  field brown dwarfs. Our results suggest the same is true at low
  gravity, and that model predictions of the cool, low mass exoplanets to
  be discovered could be in fact more accurate than model predictions
  of the few warmer L/T transition exoplanets we already know.

\begin{table*}
\caption{Colours of CFBDSIR2149 and  Ross458C.  \label{2149_458}}
\begin{tabular}{l c c c c c}
            & $Y-J$ & $J-H$ & $J-K_s$ & $H-K_s$ \\ \hline \hline
CFBDSIR2149 &  1.35$\pm$0.09   &  -0.41$\pm$0.11   &   0.13$\pm$0.09 &  0.54 $\pm$0.14      \\ \hline 
Ross458C    &  1.52$\pm$0.02   &  -0.38$\pm$0.03     & 0.08$\pm$0.03 & 0.46$\pm$0.04       \\ \hline 
\end{tabular}
\end{table*}

\section{Discussion}
 The ability to anchor the age and metallicity of CFBDSIR2149
  through its probable membership to the AB Doradus moving group
  allowed us to 
  constrain its gravity and effective temperature with atmosphere
  models and directly corroborate these conclusions through substellar evolution
  models. Comparison of the spectra to the solar metallicity
  atmosphere 
  models points towards a 650-750\,K atmosphere
  with a log\,g of 
  3.75-4.1. Using stellar evolution models of
  \citet{Baraffe.2003}, these parameters translate into a 20\,Myr to
  200\,Myr old, 2M$_{Jup}$ to 8M$_{Jup}$ free-floating planet. These
values are therefore fully compatible with 
the age range and properties of the AB Doradus moving group, even
though ABDMG membership strongly favours the higher mass estimate
and hint, in qualitative agreement with the claim of
\citet{Burningham.2011b} that BT-Settl models underestimate the
$K$-band flux of late T dwarfs, resulting in slightly underestimating the
gravity, masses and ages of the observed T dwarfs (see Table \ref{2149_458}). However our age and
metallicity constraints do confirm that CFBDSIR2149 is a low gravity, low
mass object, though not as extreme as the BT-Settl models alone would have
implied. If its ABDMG membership is confirmed, this object
would be a 4 to 7 M$_{Jup}$ T-type free-floating planet. \\

The fact the spectrum of this probable intermediate age free-floating planet
is relatively well modelled by standard atmosphere models -with no need
of ad-hoc injection of thick clouds- hints that models
would be more accurate for the cool ($\sim700$\,K), lower mass, late-T
exoplanets than for the warmer, L/T transition exoplanets such
as HR8799bcde ($\sim900-1400$\,K). If CFBDSIR2149 ABDMG membership is
confirmed, this would show that at temperatures cooler than the L/T
transition, the overall NIR spectrum of late T objects does show low
gravity features but is not dramatically different to a field late T spectrum.
\citet{Bowler.2012} and \citet{Wahhaj.2011}
 studied a L0 (1RXS J235133.3+312720B) and a L4 (CD-35
2722B) brown dwarf companions to AB Doradus members and similarly 
concluded, for temperatures higher than the L/T transition this time,
that these objects had clear low-gravity features but that
their overall spectral energy distribution was not dramatically
different from field early L dwarfs. 
 The fact that by the age of AB Doradus objects outside of the L/T
 transition already exhibit a spectral energy distribution that is
 close to that of field brown dwarfs and standard model predictions
 should comfort the detection capability hypotheses used to design the
 upcoming SPHERE and GPI planet imager instruments, at least for such
 intermediate ages.
 \\

The comparison of CFBDSIR2149 with Ross458C, has shown 
both are
  probably young planetary mass objects. Since this analysis strongly
  favours the young age hypothesis of Ross458C, we can expect some kind of
  cascade effect leading to a downward revision of the age (and
  therefore mass) estimates of a fraction of the late T dwarfs
  population. Indeed, even 
  if CFBDSIR2149 and Ross458C are probably the youngest T dwarfs
  currently identified, their gravity sensitive features do not
  widely differ from some other known field T dwarfs. 
Many discovered late T dwarfs \citep[see for instance
][]{Knapp.2004,Burgasser.2006,Lucas.2010,Burningham.2011a,
Liu.2011} have long been classified with a low
gravity (log\,g $\leq$4.5). Since these estimates were only grounded
on the comparison of their spectra and colours to atmosphere models
that have not been observationally constrained in this temperature and
gravity range, the resulting planetary mass estimates have been seen
as unprobable lower limits. Our results hint that many of these
low-gravity late T dwarfs would actually reside below the deuterium burning mass.
 This supports the
  hypothesis that a small but significant fraction of the
  known population of late T dwarfs are free-floating planets, perhaps
  the visible counterpart of the free-floating planets population
  detected by microlensing by \citet{Sumi.2011,Strigari.2012}.\\

 The conclusions for stellar and planetary formation models would also
 be far reaching. Either the planetary mass-field brown dwarfs are mostly the
 result of a stellar formation process, which would confirm the
 fragmentation of a molecular cloud can routinely form
 objects as light as a few Jupiter masses, either these objects are mostly
 ejected planets. In this case, given that massive planets are less easily
 ejected from their original stellar system than lower mass ones, and
 that lighter planets are much more common than heavier ones
 \citep[see ][for instance]{Bonfils.2011}, this would mean
 free-floating, frozen-down versions of Jupiters, Neptunes and perhaps
 Earths are common throughout the Milky Way interstellar ranges.

However, these speculations need the confirmation of
CFBDSIR2149 as a member of ABDMG moving group to become robust
hypotheses. Our ongoing parallax measurement program for CFBDSIR2149
will improve the proper motion  measurement and  determine its precise
distance, which should ascertain whether or not it is an
ABDMG member. It will also bring strong constraints on the
absolute flux and on the radius of CFBDSIR2149, enabling to use it as
a well-characterised benchmark for young late T dwarfs and the
T-type, Jupiter-masses exoplanets that are likely to be discovered by
the upcoming SPHERE, GPI and HiCIAO instruments.


\begin{acknowledgements}
Based on observations obtained with MegaPrime/MegaCam, a joint project
of CFHT and CEA/DAPNIA, at the Canada-France-Hawaii Telescope (CFHT)
which is operated by the National Research Council (NRC) of Canada,
the Institut National des Science de l'Univers of the Centre National
de la Recherche Scientifique (CNRS) of France, and the University of
Hawaii. This work is based in part on data products produced at
TERAPIX and the Canadian Astronomy Data Centre as part of the
Canada-France-Hawaii Telescope Legacy Survey, a collaborative project
of NRC and CNRS.  
''This research has made use of the NASA/ IPAC Infrared Science
Archive, which is operated by the Jet Propulsion Laboratory,
California Institute of Technology, under contract with the National
Aeronautics and Space Administration." 
We acknowledge financial support from ''Programme National de Physique Stellaire" (PNPS) of CNRS/INSU, France
\end{acknowledgements}

\bibliographystyle{aa}
\bibliography{biball}

\end{document}